\documentclass[journal]{IEEEtran}

\usepackage[font=small,skip=10pt]{caption}
\usepackage{cite}
\usepackage{caption,subcaption}
\usepackage{dblfloatfix}
\usepackage{todonotes}
\usepackage{RobStd}
\usepackage{multirow}
\usepackage{paralist}
\usepackage{cprotect}
\usepackage{amssymb}
\usepackage{dirtytalk}
\usepackage{amsmath}
\usepackage{times}
\usepackage{tikz}
\usetikzlibrary{positioning}
\usepackage{pgfplotstable}
\usepgfplotslibrary{fillbetween}
\usetikzlibrary{patterns}
\usepackage{subcaption}
\usepackage{subfiles}
\usepackage{xcolor}
\usepackage{caption}
\usepackage{cite}
\usepackage{amsmath,amssymb,amsfonts}
\usepackage{graphicx}
\usepackage{textcomp}
\usepackage{xcolor}
\usepackage{booktabs} 
\usepackage{array}
\usepackage{threeparttable}
\usepackage{cleveref}[2012/02/15]

\usepackage[noend]{algpseudocode}

\graphicspath{{images/}}

\usepackage{mathtools}
\DeclarePairedDelimiter\floor{\lfloor}{\rfloor}

\begin{document}

\title{Computing-in-Memory for Performance and Energy Efficient Homomorphic Encryption}


\newcommand{\bluHL}[1]{\textcolor{black}{#1}}
\crefformat{footnote}{#2\footnotemark[#1]#3}


\author{Dayane~Reis,~\IEEEmembership{Student Member,~IEEE,}
        Jonathan~Takeshita,
        Taeho~Jung,~\IEEEmembership{Member,~IEEE,}
        Michael~Niemier,~\IEEEmembership{Senior Member,~IEEE}
        and~Xiaobo~Sharon~Hu,~\IEEEmembership{Fellow,~IEEE}
\thanks{This work was supported in part by ASCENT, one of six centers in JUMP, a Semiconductor Research Corporation (SRC) program sponsored by DARPA. Date of publication xx xx, xxxx; date of current version xx xx, xxxx. (\textit{Corresponding author: Xiaobo Sharon Hu.})}
\thanks{D. Reis, J. Takeshita, T. Jung, M. Niemier, and X. S. Hu are with the Department
of Computer Science and Engineering, University of Notre Dame, Notre Dame,
IN, 46556, USA. e-mail: shu@nd.edu.}

}

\maketitle

\begin{abstract}
 Homomorphic encryption (HE) 
 allows direct computations on encrypted data. Despite numerous research efforts, the practicality of HE schemes remains to be demonstrated. In this regard, the enormous size of ciphertexts involved in HE computations degrades computational efficiency. Near-memory Processing (NMP) and Computing-in-memory (CiM) --- paradigms where computation is done within the memory boundaries --- represent architectural solutions for reducing latency and energy associated with data transfers in data-intensive applications such as HE. 
This paper introduces CiM-HE, a Computing-in-memory (CiM) architecture that can support operations for the B/FV scheme, a somewhat homomorphic encryption scheme for general computation. CiM-HE hardware consists of customized peripherals such as sense amplifiers, adders, bit-shifters, and sequencing circuits. The peripherals are based on CMOS technology, and could support computations with memory cells of different technologies. Circuit-level simulations are used to evaluate our CiM-HE framework assuming a 6T-SRAM memory. We compare our CiM-HE implementation 
against (i) two optimized CPU HE implementations, and (ii) an FPGA-based HE accelerator implementation.
  When compared to a CPU solution, CiM-HE obtains speedups between  4.6x and 9.1x, and energy savings between 266.4x and 532.8x for homomorphic multiplications (the most expensive HE operation). Also, a set of four end-to-end tasks, i.e., mean, variance, linear regression, and inference are up to 1.1x, 7.7x, 7.1x, and 7.5x faster (and 301.1x, 404.6x, 532.3x, and 532.8x more energy efficient). Compared to CPU-based HE in a previous work, CiM-HE obtain 14.3x speed-up and $>$2600x energy savings. Finally, our design offers 2.2x speed-up with 88.1x energy savings compared to a state-of-the-art FPGA-based accelerator.

\end{abstract}


\vspace{-5pt}
\section{Introduction}
\label{sec:introduction}

There are growing concerns regarding the security and privacy of clients' data stored in the cloud. 
Fully Homomorphic Encryption (FHE) \cite{gentry09_FHE, ducas15_fhew} may be a suitable solution for data security in the scenarios of data/computation outsourcing, because it enables any number of additions and multiplications on ciphertexts directly. However, additions and multiplications on ciphertexts accumulate noise, and a time-consuming bootstrapping mechanism is required to reduce the noises so that the final decryption contains the correct result.

Somewhat Homomorphic Encryption (SHE) \cite{fan12_BFV,bos2013improved} enables a limited number of additions and multiplications on encrypted data without losing the ability to decrypt the results of encrypted computation. Notably, multiplications cause the noise of ciphertexts to grow at a much faster pace than additions. For this reason, the depth of a SHE implementation is defined in terms of the number of multiplications it can support without invoking bootstrapping, which is denoted as \textit{multiplicative depth}. The multiplicative depth in the arithmetic circuits is limited to a constant number in SHE, however, as highlighted in \cite{roy19_hpca},  the number of operations on encrypted data is finite in many real-life applications. Therefore, the application of SHE instead of FHE may be sufficient and more practical because computationally intensive bootstrapping can be avoided. 

While SHE enables secure and private computing in the cloud, 
the large size of ciphertexts (hundreds of KB) 
may result in low speeds due to long computation times as well as the need to constantly transfer data between memory and the CPU. New computing paradigms such as Near-Memory Processing (NMP) and Computing-in-Memory (CiM) are potential solutions for reducing the overhead of data transfers between memory and the CPU \cite{jeloka16, pawlowski2011hybrid, ahn15_graph, ahn15_pimenabled, glova19}. NMP reduces the energy and latency associated with memory accesses by placing processing units close to the memory. Alternatively, CiM lowers the number of overall memory accesses by integrating certain logic and arithmetic operations directly in either the memory cells or memory peripherals. Compounded with the benefit of fewer data transfers, CiM may offer further speedups in computation due to the inherently high internal bandwidth of the memory \cite{jain17}. In other words, CiM-based solutions enable a high level of parallelism in their operations by processing many words simultaneously without data movement \cite{jain17}, which could reduce the time overheads in HE.


Recent research efforts have investigated whether compute- and data-intensive applications can be accelerated with NMP and CiM architectures based on different technologies, e.g., in \cite{aga17, jeloka16, imani19_floatpim, chi19_prime, feinberg_memristive,jain17,li16,reis18,laguna19_fewshot, gupta19}. However, most of the previous work on CiM/NMP targets general purpose computation or specific machine learning tasks. When compared to traditional applications, HE \textbf{(i)} has much longer data words --- in the order of hundreds of bits, and \textbf{(ii)} features previously unsupported operations such as modulo reduction and scaling (division with rounding). That said, the CiM infrastructure in previous works have not been designed to perform multiplication of long words and polynomials efficiently (as required by some HE schemes). Furthermore, the integer $X$ modulo $q=2^{k}$ operation and the \textit{PolyScale} primitive present in HE require conditional execution of steps by the CiM components, as well as arbitrary division and multiplication by powers of two, which cannot be supported in previous works.

In this work, we introduce a CiM architecture to support the processing of encrypted data by employing the well-known Brakerski/Fan-Vercauteren (B/FV) SHE scheme that is designed with polynomial rings\cite{fan12_BFV}. While we focus on the B/FV cryptosystem in this work, our architecture is directly applicable to other encryption schemes that incorporate similar operations (e.g. the BGV \cite{brakerski2014leveled}, GSW \cite{GSW}, TFHE \cite{TFHE}, and CKKS \cite{cheon17_CKKS} schemes\bluHL{)}. Our proposed CiM-HE targets four primitive polynomial operations used in the B/FV scheme: polynomial addition (\textit{PolyAdd}), polynomial subtraction (\textit{PolySub}), rounded polynomial scaling (\textit{PolyScale}), and polynomial multiplication (\textit{PolyMult}). Furthermore, we propose restrictions on the parameter settings to facilitate CiM-friendly solutions. Namely, CiM-HE employs moduli of $2^k$ for some integer $k$. Also, the \textit{PolyScale} primitive assumes a divisor of the form $2^{k'}$ for some integer $k'$. Our parameter settings do not impact the security of HE, but rather make it CiM-friendly. \bluHL{Below, we list the main contributions of our work:}
\begin{itemize}
    \item \bluHL{We are the first to support the execution of all essential evaluation operations of the B/FV SHE scheme within a CiM framework. CiM-HE proposes a mapping of polynomial primitives to the underlying CiM-HE hardware, with support to \textit{polynomial ring operations};}
    \item \bluHL{We implement bit shifts of variable lengths (accomplished with an in-memory logarithmic shifter), conditional execution in algorithms (through flag generation in our sequencing circuit), and coefficient re-placement in memory (with in-place move buffers). These novel CiM capabilities are introduced in this work for the purpose of supporting the aforementioned polynomial primitives;}
    \item \bluHL{We provide real end-to-end use case studies (and a thorough evaluation) for a SHE scheme running on our CiM-HE architecture without the use of bootstrapping, i.e., with a multiplicative depth of 5.}
\end{itemize}

We perform simulations using Verilog and SPICE to evaluate speed and energy consumption of CiM-HE. We employ the BSIM-CMG FinFET model from \cite{duarte15_bsim_cmg} for a 14nm technology node, and compare the runtime of three HE primitives (\textit{HomAdd}, \textit{HomSub}, and \textit{HomMult}, based on the four primitive polynomial operations) in our CiM-HE design --- without algorithmic optimizations --- against a CPU-based implementation based on the widely-used Microsoft Simple Encrypted Arithmetic Library (SEAL) \cite{chen17_SEAL} with various algorithmic optimizations enabled. Our results indicate speedups of $>$10000x (4.6x) for homomorphic addition (multiplication), and energy savings for addition (multiplication) of $>$290x ($>$530x) for a single execution of each homomorphic operation.

We also demonstrate the benefits of CiM-HE by evaluating widely used operations in data analytics, i.e., arithmetic mean, variance and linear regression, as well as inference, on encrypted data using HE primitives. The speedups (and energy savings) of CiM-HE varies by task. In our experiments, we obtain maximum (minimum) speedups of $>$26000x (1.1x) and maximum (minimum) energy savings of 532.8x (299.7x). Better speedups are obtained for tasks whose execution time is dominated by additions, e.g., in the arithmetic mean. Furthermore, compared to an alternative CPU-based HE based from \cite{bos2017privacy}, we obtain a 14.3x speedup and $>$2600x energy savings for homomorphic multiplications. Finally, compared to a state-of-the-art FPGA-based SHE accelerator \cite{roy19_hpca},  our design is 88.1x more energy efficient with speedup of 2.2x for a security level of 80 bits and a multiplicative depth of 4.

\section{Background and Preliminaries}
\label{sec:background}

\subsection {B/FV scheme based on RLWE and its polynomial primitives}
\label{sec:fhe_introduction}


SHE schemes based on polynomial rings and the Ring Learning-With-Error (RLWE) problem \cite{fan12_BFV,brakerski2013classical} have become popular due to their security against quantum algorithms. Their implementations are widely available and actively maintained as well \cite{halevi2013design,chen17_SEAL,TFHE}. 
\bluHL{We focus on the well-known B/FV scheme \cite{fan12_BFV} in this paper. Our design is easily extensible to other lattice-based cryptographic schemes, including the BGV \cite{brakerski2014leveled}, GSW \cite{GSW}, TFHE \cite{TFHE}, and CKKS \cite{cheon17_CKKS} schemes. Because all of these schemes rely on the same low-level operations (i.e., modular polynomial arithmetic such as polynomial addition and multiplication), we can apply our design to these schemes by choosing the sequences of primitive operations appropriately, as we do in Section \ref{sec:hardware_mapping} for B/FV.} 

The B/FV scheme takes in plaintext messages from $\mathcal{R}_t:=\mathbb{Z}_t[X]/\Phi_M(X)$, i.e., the set of residue classes of the polynomial $\Phi_M(X)$ whose coefficients are from $\mathbb{Z}_t$, and operates on ciphertext polynomials in $\mathcal{R}_q:=\mathbb{Z}_q[X]/\Phi_M(X)$. These sets form rings of polynomials under polynomial addition and multiplication. 
We provide a brief overview of B/FV SHE below. Details with respect to key generation, encryption, and decryption are omitted as the focus of this work is on the acceleration of homomorphic operations on encrypted data\footnote{In the applications of homomorphic encryption, key generation, encryption and decryption are one-time processes, and the bottleneck lies in the homomorphic operations.  The details can be found in the original publication \cite{fan12_BFV}.}.

A plaintext message (e.g., numbers) can be encoded into a plaintext polynomial in $\mathcal{R}_t$ with standard encoding techniques (e.g., integer encoding, fraction encoding
), which is then encrypted into a pair of polynomials in $\mathcal{R}_q^{2}$, which is the ciphertext $c=(c[0],c[1])$ of the plaintext polynomial.

Given two ciphertexts $c_{1}$ and $c_{2}$ which are ciphertexts of two messages $m_1,m_2$ respectively, homomorphic addition is defined as:
\begin{equation}
HomAdd(c_1,c_2)=([c_1[0]+c_2[0]]_q,[c_1[1]+c_2[1]]_q)
\end{equation}
where $q$ is the moduli of ciphertext polynomials and $[\cdot]_q$ for an integer denotes the reduction modulo $q$ defined as $[x]_q=x-\lfloor x/q\rceil\cdot q$. When the input of $[\cdot]_q$ is a polynomial, the reduction modulo $q$ is applied coefficientwise. The resulting ciphertext is the encryption of $m_1+m_2$.

In Microsoft's SEAL \cite{chen17_SEAL} that implements the B/FV scheme,  homomorphic subtraction between two ciphertexts $c_{1}$ (minuend) and $c_{2}$ (subtrahend) is  implemented as:
\begin{equation}
HomSub(c_1,c_2)=([c_1[0]-c_2[0]]_q,[c_1[1]-c_2[1]]_q)
\end{equation}
Similarly, the resulting ciphertext is the encrypted subtraction $m_1-m_2$.

The homomorphic multiplication between two ciphertexts $c_{1}$ and $c_{2}$ is defined as a series of polynomial operations.
First, the following polynomials are computed:
\begin{equation}
\begin{split}
    c_x&=[\lfloor{t(c_1[0]\cdot c_2[0])}/{q}\rceil]_q\\
    c_y&=[\lfloor{t(c_1[0]\cdot c_2[1]+c_1[1]\cdot c_2[0])}/{q}\rceil]_q\\
    c_z&=[\lfloor{t(c_1[1]\cdot c_2[1])}/{q}\rceil]_q\\
\end{split}
\end{equation}
Here, $t$ is the moduli of plaintext polynomials.
Then, the \textit{relinearization key} $rlk=(rlk_0,rlk_1)$ generated from the key generation algorithm (which is a pair of polynomials) is used to generate the final ciphertext, also a pair of polynomials:
\begin{equation}
\begin{split}
    HomMul(c_1,c_2)&=(c_x+\langle rlk_0,c_z \rangle,c_y+\langle rlk_1,c_z\rangle)
\end{split}
\end{equation}
Here, $\langle poly_x,poly_y\rangle$ represents the inner product between two polynomials (i.e., the sum of coefficient-wise multiplications). The resulting ciphertext is the encrypted multiplication $m_1\cdot m_2$.

The computations above are comprised of two types of operations:  {\bf (i) ring operations}, which are polynomial additions, subtractions and multiplications. These consist of modular additions and multiplications closed in $\mathbb{Z}_q$. For ring operations, we support three primitive operations on polynomials: polynomial addition (\textit{PolyAdd}), polynomial subtraction (\textit{PolySub}), and polynomial multiplication (\textit{PolyMult}). Additionally, {\bf (ii) rounded polynomial scaling} scales a polynomial by $t/q$ and returns the closest polynomial in the ring, i.e., coefficients are rounded to the nearest integer and the tie is broken by rounding it up. 
For this step, our CiM-HE implements the rounded polynomial scaling (\textit{PolyScale}) as the final primitive.

\subsubsection{Bootstrapping}

\bluHL{Fully homomorphic encryption schemes can achieve computation of arbitrary circuits not limited by multiplicative depth with a procedure called bootstrapping. Random noise accumulates in ciphertexts of lattice-based homomorphic encryption with each operation. After a number of multiplications determined by a scheme's multiplicative depth, further accumulated noise makes correct decryption impossible. At that point, the ciphertext must either be decrypted, or subjected to computationally expensive bootstrapping to remove the noise. Due to the complexity of bootstrapping, implementations of homomorphic encryption usually only implement the SHE portions of the scheme and do not include bootstrapping \cite{chen17_SEAL,riazi2020heax,roy19_hpca,cousins17_fpga_accelerator,bajard2016full,halevi2019improved,poppelmann2015accelerating}, making the implementation leveled, i.e., supporting arbitrary circuits up to certain multiplicative depth. We also only implement SHE operations and do not include bootstrapping. Although we do not explicitly present how bootstrapping is implemented, our architecture can easily support the bootstrapping operations of B/FV, BGV, CKKS. These operations boil down to homomorphic evaluation of certain arithmetic circuits (i.e., a sequence of homomorphic additions, subtractions, and multiplications) over the bootstrapped ciphertexts \cite{cheon2018bootstrapping,halevi2015bootstrapping,fan12_BFV}, therefore they can be supported by our architecture which implements homomorphic addition, subtraction, and multiplication.  Notably, with our choices of parameters (outlined in Section \ref{sec:parameter_settings}), we are even able to take advantage of a special optimized case of bootstrapping when the ciphertext and plaintext moduli are powers of two \cite{fan12_BFV}.}

\subsubsection{Rotation}

\bluHL{Some RLWE-based schemes (most notably CKKS) allow batched encryption of many plaintexts into a single ciphertext, which can then be operated upon in a SIMD fashion \cite{cheon17_CKKS}. Rotations and permutations on the order of the batched plaintexts can then be performed homomorphically upon the ciphertext. Homomorphic rotation and permutation are not core SHE operations for the B/FV scheme, so we do not evaluate those in this work. We note that these operations can be done efficiently using CiM (with appropriate parameter choices), as they only involve a polynomial rotation/permutation and simple arithmetic (the polynomial addition, multiplication, and division-with-rounding needed to compute a key switch). Specifically, given a CKKS ciphertext $(c[0], c[1])$ at a level $\ell$, the computation required is to find $(\pi(c[0]), 0) + \lfloor P^{-1} \cdot \pi(c[1]) \cdot \textbf{swk} \rceil \mod q_\ell$, where $\pi$ is a permutation on the polynomial's coefficients, $\textbf{swk}$ is a special key, and $P^{-1}, q_\ell$ are parameters of CKKS \cite{cheon2018bootstrapping}.}

\subsection{Related Work}
\label{sec:related_work}

\subsubsection{CiM and NMP designs for HE}
CiM enables significant speedups and energy savings for data-intensive problems in multiple application domains, e.g., \cite{jeloka16, pawlowski2011hybrid, ahn15_graph, ahn15_pimenabled, glova19, imani19_floatpim, chi19_prime, feinberg_memristive,jain17,li16,reis18, laguna19_fewshot}. CiM designs targeting SRAM-based caches \cite{jeloka16} or main memory \cite{pawlowski2011hybrid} have been fabricated. 
With respect to security-centric applications, CiM-based engines have been used for realizing encryption schemes such as AES \cite{wang14,xie18}. However, there is limited work regarding ``in-memory" HE due to the complexity of operations involved in these encryption schemes (e.g., multiplication between ciphertexts that are hundreds of KB in size). In this regard, SCAM \cite{bian17_SCAM} proposes a modification to the homomorphic XOR and AND operations as defined in the FHEW scheme \cite{ducas2015fhew} in order to perform search operations on encrypted data using a fully-additive search function. CiM and NMP implementations of this search function were presented in \cite{reis19} and \cite{glova19}, respectively. CiM and NMP approaches enable speedups and energy savings compared to \cite{bian17_SCAM}. To the best of our knowledge, there are no other works on the implementation of both homomorphic additions and homomorphic multiplications in either CiM or NMP.

\subsubsection{Hardware Accelerators}

Several hardware accelerators for HE have been proposed \cite{poppelmann2015accelerating,roy15_modular,wang14_tvlsi,xiaolin14,cilardo16_fpga,cousins17_fpga_accelerator,roy19_hpca}. Most designs are based on Field Programmable Gate Arrays (FPGAs), which provide an option for implementation of varied HE circuits. Some designs, e.g., \cite{wang14_tvlsi,xiaolin14,cilardo16_fpga}, focus on the design of multiplier units for very large numbers, which are useful for performing homomorphic multiplications. 

Accelerators for general HE computation are proposed in \cite{poppelmann2015accelerating,roy15_modular,cilardo16_fpga,cousins17_fpga_accelerator,roy19_hpca}. References \cite{poppelmann2015accelerating,roy15_modular} and \cite{riazi2020heax} propose accelerating the YASHE \cite{bos2013improved} and CKKS schemes, respectively. The YASHE scheme has prohibitively large ciphertext sizes, which aggravates the impact of memory transfers on performance due to memory bandwidth limitations. This fact is not accounted for in \cite{roy15_modular}. Furthermore, the YASHE scheme is proven insecure as it is subject to a subfield lattice attack \cite{albrecht2016subfield}. As for CKKS, this scheme is also based on polynomials operations (like the B/FV), but it can operate on floating-point numbers, which is amenable to machine learning applications. \bluHL{In \cite{riazi2020heax}, an FPGA-based accelerator provides support for all the essential HE operations in the CKKS scheme. Improvements of two orders of magnitude in homomorphic multiplication time are reported when compared to a CPU solution that implements CKKS. However, the work does not provide an evaluation for a larger workload, i.e., end-to-end tasks applying many different operations on tens or hundreds of ciphertexts as likely to occur in a real-life setting are not included in the evaluation.}   

In this work we evaluate the functionality of CiM-HE for the B/FV scheme. Likewise, \cite{cousins17_fpga_accelerator, roy19_hpca} can perform the essential operations of the B/FV SHE scheme, i.e., homomorphic multiplications and additions. To ensure the fairest comparison possible between different works, we do not compare CiM-HE to others that are based on different schemes, e.g., YASHE, CKKS, etc. Regarding prior work that focuses on the B/FV scheme, reference \cite{cousins17_fpga_accelerator} reports a homomorphic multiplication time of 463.9 ms for 80-bit security, which is $\sim$92.8x more than the time reported in \cite{roy19_hpca} for the same operation with same security level. As \cite{roy19_hpca} represents a state-of-the-art FPGA-based HE accelerator for the B/FV scheme, we choose to compare CiM-HE with that work in our evaluation (see Sec. \ref{sec:evaluation}).

\section{CiM-HE Framework}
\label{sec:CiM-HE_framework}

In this section, we describe our proposed CiM-HE framework, which consists of CiM hardware as well as the mapping of primitives (\textit{PolyAdd}, \textit{PolySub}, \textit{PolyScale}, and \textit{PolyMult})
to a sequence of CiM-friendly steps. We also discuss HE parameter settings used by the CiM-HE framework.

\subsection{Parameter settings for CiM-friendly HE}
\label{sec:parameter_settings}

Per Sec. \ref{sec:background}, polynomial primitives are always computed over polynomial rings $\mathcal{R}_q$ with moduli $q$ for coefficients. To expedite the integer reduction modulo $q$ in memory, we set $q$ to be a power of two. We also set the divisor (for the \textit{PolyScale}) to be a power of two. Thus, the polynomial ring of interest is $\mathcal{R}_{2^k}:=\mathbb{Z}_{2^k}[X]/\Phi(X)$ for some positive integer $k$. When divisions are needed during the polynomial scaling, we divide the coefficients by powers of two, which can be done by performing bit shifting in memory. This choice of parameters \textbf{does not affect the security of HE schemes}\cite{fan12_BFV,langlois2012hardness,peikert2009public,regev2009lattices}, because the RLWE problem is hard regardless of whether the moduli of the polynomial coefficients are powers of two instead of prime numbers or products of coprime numbers.

We have followed the widely-accepted analysis of Lindner and Peikert \cite{lindner2011better} to choose the parameters $n$ (ciphertext polynomial degree) and $log_2(q)$ (ciphertext modulus size) to achieve an acceptable level of security in CiM-HE, i.e., $\geq 128$ bits. In Sec. \ref{sec:evaluation}, we compare the performance and energy consumption of CiM-HE to a CPU-based implementation with four [$n$, $log_2(q)$] pairs to achieve different multiplicative depths and levels of security. Furthermore, in Sec. \ref{sec:evaluation}, we also compare CiM-HE with previous work, assuming the same parameter setting as \cite{roy19_hpca,bos2017privacy}, which enables us to achieve 80-bit security and a multiplicative depth of 4.


\begin{figure}[!t]
    \centering
    \includegraphics[width=1\columnwidth]{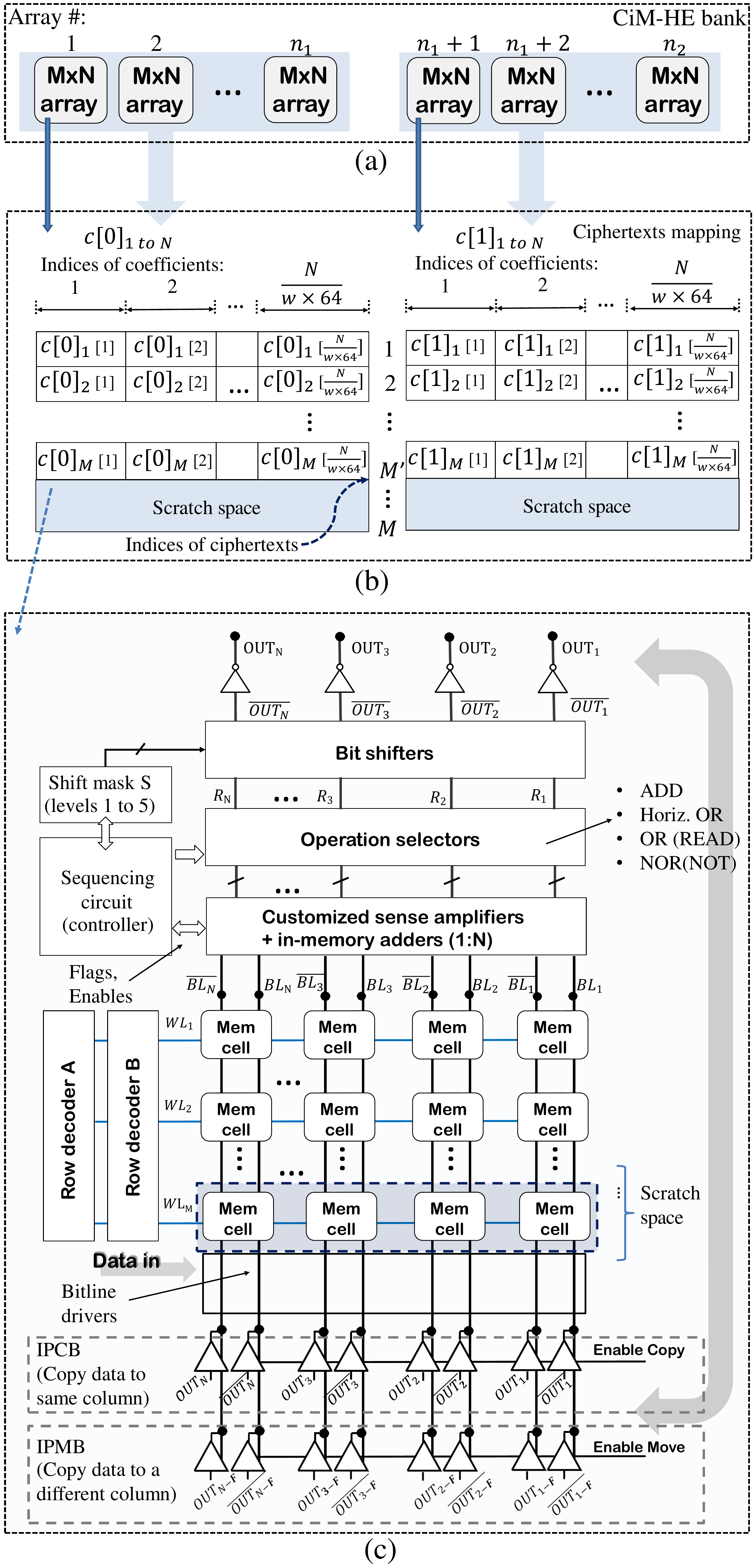}
    \caption{(a) A CiM-HE bank, where (b) mapping of coefficient to CiM-HE bank and (c) individual arrays are highlighted. Each CiM array consists of memory cells and peripherals that support the execution of polynomial primitives. }
    \label{fig:cim_hardware}
\end{figure}
\begin{figure}[!t]
    \centering
    \includegraphics[width=1\columnwidth]{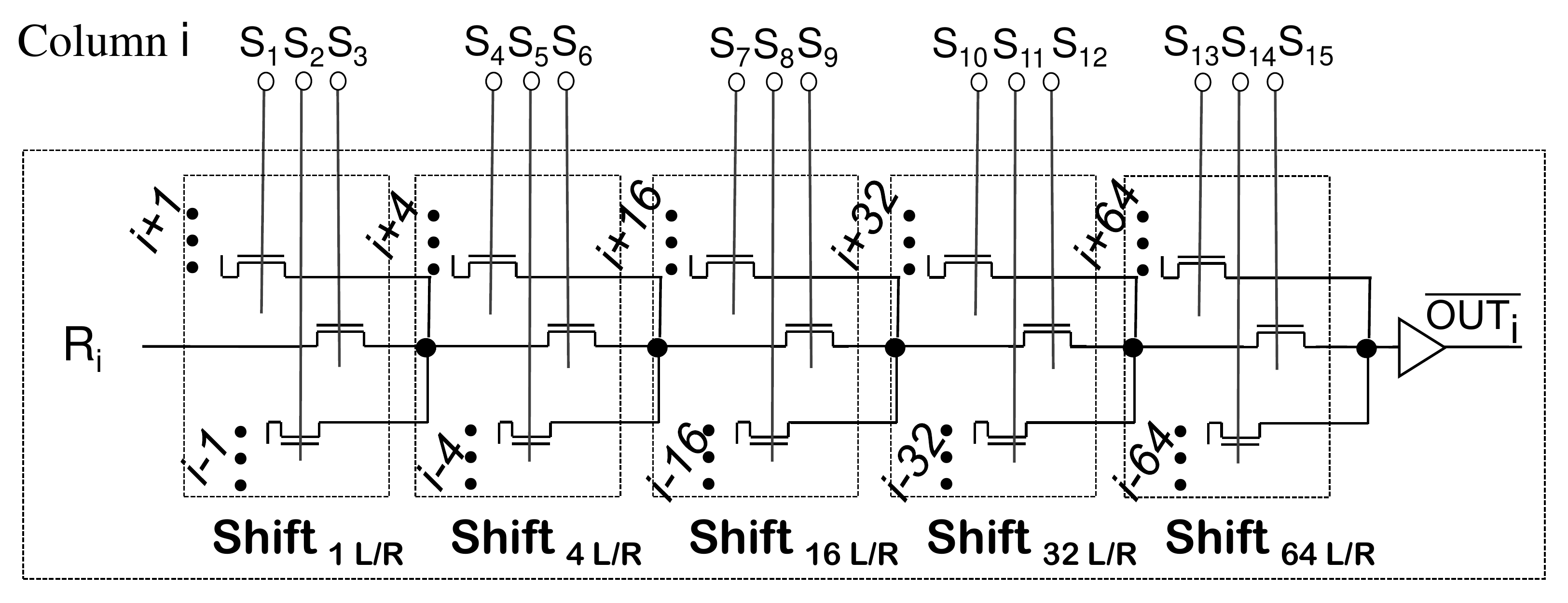}
    \caption{Log shifter implemented in CiM-HE.}
    \label{fig:cim_log_shifter}
    \vspace{-2ex}
\end{figure}

\subsection{CiM hardware}
\label{sec:CiM_hardware_description}

Fig. \ref{fig:cim_hardware} illustrates our CiM hardware architecture.
A CiM-HE bank (Fig. \ref{fig:cim_hardware}(a)) consists of multiple CiM-HE arrays of $M \times N$ size where ciphertexts are stored. When designing a bank, we should choose the number and dimension of the arrays based on {\bf (i)} the number of ciphertexts we wish to store in a single CiM-HE bank, {\bf (ii)} the coefficient's modulus ($q=2^{k}$), and {\bf (iii)} the degree of polynomials ($n$). When positioning polynomials in the CiM-HE banks before a HE operation, we have to write the polynomials in a column-aligned fashion (so coefficients of the same degree are mapped to the same columns in the CiM arrays) to facilitate the execution of polynomial primitives by our CiM-HE hardware.

Figs. \ref{fig:cim_hardware}(a) and \ref{fig:cim_hardware}(b) illustrate the mapping of polynomial coefficients to the memory arrays in a CiM-HE bank. Recall that the ciphertext has the form $c=(c[0],c[1])$. In  Fig. \ref{fig:cim_hardware}(a), we define arrays numbered $1$ to $n_{1}$ to store the first polynomial ($c[0]$), while arrays numbered $n_{1}+1$ to $n_{2}$ store the second polynomial ($c[1]$). The coefficient mappings depicted in Fig. \ref{fig:cim_hardware}(b) are for arrays $1$ and $n_{1}+1$. $M$ and $N$ are the number of rows and columns in an array, respectively. $M-M'$ is the number of rows reserved as a scratch space to store intermediate results of CiM-HE operations. The coefficients of the two polynomials $c[0]$ and $c[1]$ are labeled as ${c[0]_{y}}{[z]}$ and ${c[1]_{y}}{[z]}$, where $y$ is the index for the ciphertexts, and $z$ is the index for coefficients that are stored in memory. As such, each array holds \textit{up to} $C=N/(w\times 64)$ coefficients of $M'$ ciphertexts. In this equation, $w$ represents the number of standard-sized 64-bit words that are necessary to store each $k$-bit coefficient. Namely, $w\times 64\geq k$. 
The coefficients of $M$ different ciphertexts of the same degree are placed in the same column to facilitate in-memory logic and arithmetic operations such as AND, OR, addition, and subtraction.  The in-memory computation of polynomial primitives in each array is supported by a set of customized memory peripherals, including a sequencing circuit (controller), customized sense-amplifiers, in-memory adders 
and bit shifters, which are described below. 

\subsubsection{Controller}

The controller activates appropriate clock signals so the customized sense amplifier and in-memory adder circuits can perform operations at the correct times. 
In addition to operation sequencing, the controller receives flag inputs from the customized sense amplifiers that allow for the conditional execution of steps in a reduction of integer $X$ modulo $q=2^{k}$ operation and the \textit{PolyScale} primitive (see Sec. \ref{sec:poly_int_reduction} and Sec. \ref{sec:poly_scaling}).  

\subsubsection{Customized sense amplifiers/in-memory adders}

The customized sense amplifiers and in-memory adders employed in this work perform bitwise logic and wordwise addition between two ciphertexts stored in memory. The in-memory adder is also used to implement subtraction and coefficient multiplication in \textit{PolySub} and \textit{PolyMult} primitives. We employ circuits similar to those in \cite{reis19, aga17}, i.e., an in-memory carry select adder and a customized sense amplifier that enables bitwise logic between two memory words. We also introduce a new operation, ``horizontal OR", in which the bitwise AND result is ORed horizontally, i.e., within all the bits of a word. The ``horizontal OR" operation is used to generate modulo reduction and rounding flags for the execution of the \textit{PolyScale} primitive (see Sec. \ref{sec:poly_int_reduction} and Sec. \ref{sec:poly_scaling}). At the end of each computation, we keep the output (computation result) stored in a temporary latch, so it can be copied/moved to an address in memory in the subsequent cycle.

\subsubsection{Bit shifters}
\label{sec:bit_shifters}
The bit shifters implemented in CiM-HE can perform right bit shifts that enable divisions by powers of two for the  \textit{PolyScale} primitive. We choose to implement a logarithmic (log) shifter in the memory (Fig. \ref{fig:cim_log_shifter}). Compared to other bit shifters such as a barrel shifter, the log shifter employed in CiM-HE offers flexibility in terms of the number of shifts possible to achieve in a single round of computation with less area overhead. For instance, a log shifter with $L$ levels can perform $2^{L}$ combinations of right/left shift in a single round of bit shifting, and multiple rounds can be used to achieve any desired number of right/left bit shifts. It is possible to arbitrarily choose the number of bit shifts each level performs to favor either larger or smaller shifts in one cycle. For instance, Fig. \ref{fig:cim_log_shifter} depicts the log shifter in our CiM-HE. We have a 5-level log shifter; each level enables 1, 4, 16, 32, or 64 bit shifts, which can be combined in a single cycle to achieve greater numbers of shifts, e.g., ``64+32+16+4+1=117", or ``64+4+1=69". It is possible to activate only certain levels of the log shifter and keep others deactivated, i.e., performing a shift by 0. The shift mask $S_{1}-S_{15}$, a 15-bit number, is used to activate left/right or no shifts in every level $L$ of the log shifter. Note that, when setting a mask, we must make sure that one (and only one) transistor is selected per level.

\subsubsection{In-place copy buffers/In-place move buffers}
\label{sec:buffers}
In-place copy buffers (IPCBs) based on \cite{aga17} are used to copy CiM outputs (i.e., the results of a bitwise logic, addition, or right/left shift) to an address defined in the memory scratch space. IPCBs are always placed in alignment with each sense amplifier column in the memory array. As such, they do not allow data movement between column $i$ and $i+F$ ($F$ is a pre-defined offset). As CiM-HE needs both (i) a copy of outputs to same column in the array, and (ii) data movement between distinct columns for the execution of polynomial operations, we introduce in-place move buffers (IPMBs) in our architecture. While IPMBs have a similar schematic as IPCBs, we use a different routing scheme in IPMBs that allows us to move data from column $i$ to column $i+F$ in an array, where $F$ is a pre-defined column offset. Note that by performing this operation in $c$ cycles, it is possible to move data with $cF$ different offsets.

\subsection{Mapping operations in SHE to CiM hardware}
\label{sec:hardware_mapping}

We now describe how CiM-HE components can be used to execute polynomial primitives that are used to support the computational requirements of the B/FV scheme. Each primitive is broken down into a sequence of CiM-friendly steps that require one or more cycles to be executed. Note that we reduce integers $X$ modulo $q$ (as described below) after the execution of \textbf{every polynomial primitive}, to ensure that our results are in the ring $\mathcal{R}_q$.

\subsubsection{Reduction of integer $X$ modulo $q\!=\!2^{k}$}
\label{sec:poly_int_reduction}

Because all coefficients of polynomials are closed in $\mathbb{Z}_q$, the first operation we need to support is the integer reduction modulo $q$. Computing the reduction of integer $X$ modulo $q\!=\!2^{k}$  (i.e. $[x]_q\in[-\frac{q}{2},\frac{q}{2})$) after the execution of polynomial primitives. Integer reduction modulo $q\!=\!2^{k}$ in CiM-HE is facilitated by the fact that the ${k}^{th}$ bit --- and all other bits to the left --- 
represent the integer part when performing division by $q$. As such, these bits can be ignored if we are only interested in the modulo. While simple, the aforementioned method only returns moduli in the interval $[{0},{q-1}]$. To find the modulo in the correct interval, i.e. $[x]_q\in[-\frac{q}{2},\frac{q}{2})$, additional CiM steps are required. We compare half of $q$ (the bit in the $(k-1)^{th}$ position) to the value of our current modulo via a three-step process.

\begin{itemize}
    \item {\bf Step 1:} We store a mask with all `0's and a single `1' in the ${(k-1)}^{th}$ position in a scratch space of the CiM memory. We perform a bitwise in-memory AND between this mask and the residues from $[{0},{q-1}]$, which results in a $k$-bit number (an intermediate result).
    \item {\bf Step 2:} At the customized sense amplifier, we perform a horizontal OR between the ${k}$ bits of the intermediate result. The horizontal OR operation generates a flag value, either `0' or `1', which is used by the CiM controller to indicate the next operation. 
    \item {\bf Step 3}: If the flag is `0', the current modulo is a positive number in the interval $[{0},\frac{q}{2})$, which satisfies our desired form for modulo reduction. If the flag is `1', a \textit{PolySub} operation between the current modulo and $q$ has to performed, so a modulo in the interval $[-\frac{q}{2},0)$ can be found ($q$ is stored in the scratch space of the CiM memory).
\end{itemize}
    
\subsubsection{Homomorphic Addition via PolyAdd}
\label{sec:poly_addition}

Homomorphic addition (Sec. \ref{sec:fhe_introduction}) requires polynomial additions, i.e., \textit{PolyAdd}. As long as coefficients of the same order of two (or more) polynomials are placed in the memory in a column-aligned fashion, this can be directly realized by the in-memory adder described in Sec. \ref{sec:CiM_hardware_description}.

\subsubsection{Homomorphic Subtraction via PolySub}
\label{sec:poly_subtraction}

The homomorphic subtraction (Sec. \ref{sec:fhe_introduction}) requires polynomial subtractions, i.e.,
\textit{PolySub}. This can also be realized by the in-memory adder described in Sec. \ref{sec:CiM_hardware_description}, but requires an extra step before the addition itself, which consists of finding the 2's complement of the subtrahend. The \textit{PolySub} primitive can be performed as a two-step procedure.

\begin{itemize}
    \item {\bf Step 1}: Performing a NOT operation on the subtrahend with our customized sense amplifier, and copying the result to the scratchpad space in the memory array.
    \item {\bf Step 2}: Adding the minuend with inverted subtrahend, while setting the carry-in bit of the in-memory adder to `1'. (This is equivalent to performing the subtraction via 2's complement arithmetic.)
\end{itemize}

\subsubsection{\textit{PolyScale} for Homomorphic Multiplication}
\label{sec:poly_scaling}

The homomorphic multiplication (Sec. \ref{sec:fhe_introduction}) involves rounded scaling. Therefore, before introducing \textit{HomMult}, we describe how our CiM-HE architecture supports scaling.
Our CiM-HE can perform rounded scaling by powers of two. Given a polynomial $c$ and a divisor in the form $D=2^{k'}$, we perform  rounded scaling by performing a division by ${2^{k'}}$ for every coefficient, which returns the integer quotient for every coefficient. Divisions by powers of two in memory can be implemented with our CiM log-shifter circuit (described in Sec. \ref{sec:CiM_hardware_description}). 


The log shifter in CiM-HE has 5 levels that allows us to have either 0, or a pre-defined number of right shifts in each level. The pre-defined shifts in each level are 64, 32, 16, 4, and 1. Therefore, for a division by $D=2^{127}$, we break down the division into three right-shift rounds:

\begin{itemize}
    \item {\bf Round 1:} $64+32+16+4+1=117$;
    \item {\bf Round 2:} $0+0+0+4+1=5$;
    \item {\bf Round 3:} $0+0+0+4+1=5$.
\end{itemize}  

After each right-shift division round, the result is stored in the memory though IPCBs (Fig. \ref{fig:cim_hardware}(c)). To determine which CiM log-shifter levels must be activated in the next round, we perform a subtraction between $k'$ and the value of activated levels (pre-determined values that can be stored in the scratch space of the CiM array). If the result of this subtraction is negative (which can be verified by looking at its signal bit), we know that the number of bit shifts is beyond what is needed by the division. Hence, we make the highest active level equal to 0 for the next right-shift round and continue the process iteratively until we reach a number of accumulative right-shifts equal to $k'$.

After performing the division with right-shifts, we need to round the value of $z$ to the nearest integer (round up/down). This can be done by a two-step procedure as follows:

\begin{itemize}
    \item {\bf Step 1:} We compare the bits to the right of $k'$ in the dividend, (i.e., the remainder) to half of the divisor. We use a bit mask with all `0's and a single `1' in the $({k'-1})^{th}$ position. We perform an in-memory AND followed by horizontal OR operations to generate a rounding flag (similar to the reduction of integer $X$ modulo $q$). 
    \item {\bf Step 2:} If the result of the AND-OR flag is '1', rounding up is performed by adding '1' to the quotient $z$. Otherwise, $z$ should be rounded down (the bit shifts perform this type of rounding by default).
\end{itemize}

\subsubsection{Homomorphic Multiplication via all polynomial primitives}
\label{sec:poly_multiplications}

\begin{algorithm}[t]
\caption{Karatsuba algorithm \cite{karatsuba1962multiplication,cesari1996performance}}
\label{alg:karatsuba_mult}
    \textbf{Input:} Polynomials $A,B$ of degree $n$ and $k$-bit coefficients\\
    \textbf{Output:} $A \times B$\\
    \textsc{Karatsuba}($A,B$)
\begin{algorithmic}[1]

    \If{$n==1$}
        \State \textbf{return} \textsc{Shift-add}($A,B$)
    \EndIf
    \State $HighA$, $LowA:=$  \textsc{Split}($A$,$\lfloor n/2 \rfloor$); \Comment{IPMBs}
    \State $HighB$, $LowB:=$  \textsc{Split}($B$,$\lfloor n/2 \rfloor$); \Comment{IPMBs}
    \State $R1:=$\textsc{Karatsuba}($HighA+LowA$,$HighB+LowB$);
    \State $R2:=$\textsc{Karatsuba}($HighA$,$HighB$); 
    \State $R3:=$\textsc{Karatsuba}($LowA$,$LowB$);
    \State $R1:=R1-R2-R3$;
    \State \textbf{return} $R2 \times 2^{2nk} + R1  \times 2^{nk} + R3$ \Comment{shifts/additions}
\end{algorithmic}
\end{algorithm}

\begin{algorithm}[t]
\caption{Shift-Add algorithm}
\label{alg:shiftadd_mult}
    \textbf{Input:} Multiplicand $a$ and multiplier $b$, which are $k$-bit coefficients\\
    \textbf{Output:} $out=a \times b$\\
    \textsc{Shift-add}($a,b$)
\begin{algorithmic}[1]
    \State $out:=0$;\Comment{reserve in memory}
    \State $a':=a$; \Comment{copy to memory}
    \State $b':=b$; \Comment{copy to controller}
    \For{$i=1:k$}
        \If{$b'(i)==0$}
            \State $a':=$\textsc{Shift}($a'$); \Comment{1-bit left shift}
        \EndIf
        \If{$b'(i)==1$}
            \State $out:=$\textsc{Add}($out,a'$);
        \EndIf
    \EndFor
    \textbf{return} $out$
\end{algorithmic}
\end{algorithm}

Homomorphic multiplication (Sec. \ref{sec:fhe_introduction}) involves multiple polynomial primitives. For example, to calculate $c_y$, one needs to perform two multiplications of two polynomials (\textit{PolyMult}), add the two resulting polynomials (\textit{PolyAdd}), and then perform rounded scaling for the resulting polynomial (\textit{PolyScale}) to round the polynomial after scaling it by $t/q$.

Implementing \textit{PolyMult} in the memory using the naive ``schoolbook" multiplication method would require $\mathcal{O}(n^{2})$ multiplications between each pair of coefficients, which could trigger impractically high data movement between CiM arrays. Thus, we employ the Karatsuba multiplication algorithm \cite{karatsuba1962multiplication}, and execute it with our CiM design. The Karatsuba algorithm recursively breaks a multiplication of two polynomials into multiplications of polynomials with half the number of terms. Compared to a ``schoolbook" method, the data movement between arrays --- and thus the complexity of implementation --- is significantly lower with Karatsuba multiplication. The steps for the Karatsuba multiplication with CiM are described in Algorithm \ref{alg:karatsuba_mult}.

The CiM implementation of the Karatsuba algorithm is executed recursively. The base case consists of a multiplication between two coefficients (in our case, $k$-bit numbers), i.e., a product that can be computed by \textit{Shift-Add} operations (Algorithm \ref{alg:shiftadd_mult}). To perform these operations in memory, we initially (i) reserve a memory address to store the accumulation of partial products ($out$), (ii) copy the value of multiplicand to the scratch space in memory (so we can shift it in each iteration), and (iii) store the value of the multiplier in a temporary register in the controller ($b'$). We access the bits of $b'$ in the $k$ subsequent cycles in order to determine whether we shift $a'$, i.e. the copy of the multiplicand, 1 bit to the left, or if we add the contents of $out$ to $a'$.

\begin{figure}[!t]
    \centering
    \includegraphics[width=1\columnwidth]{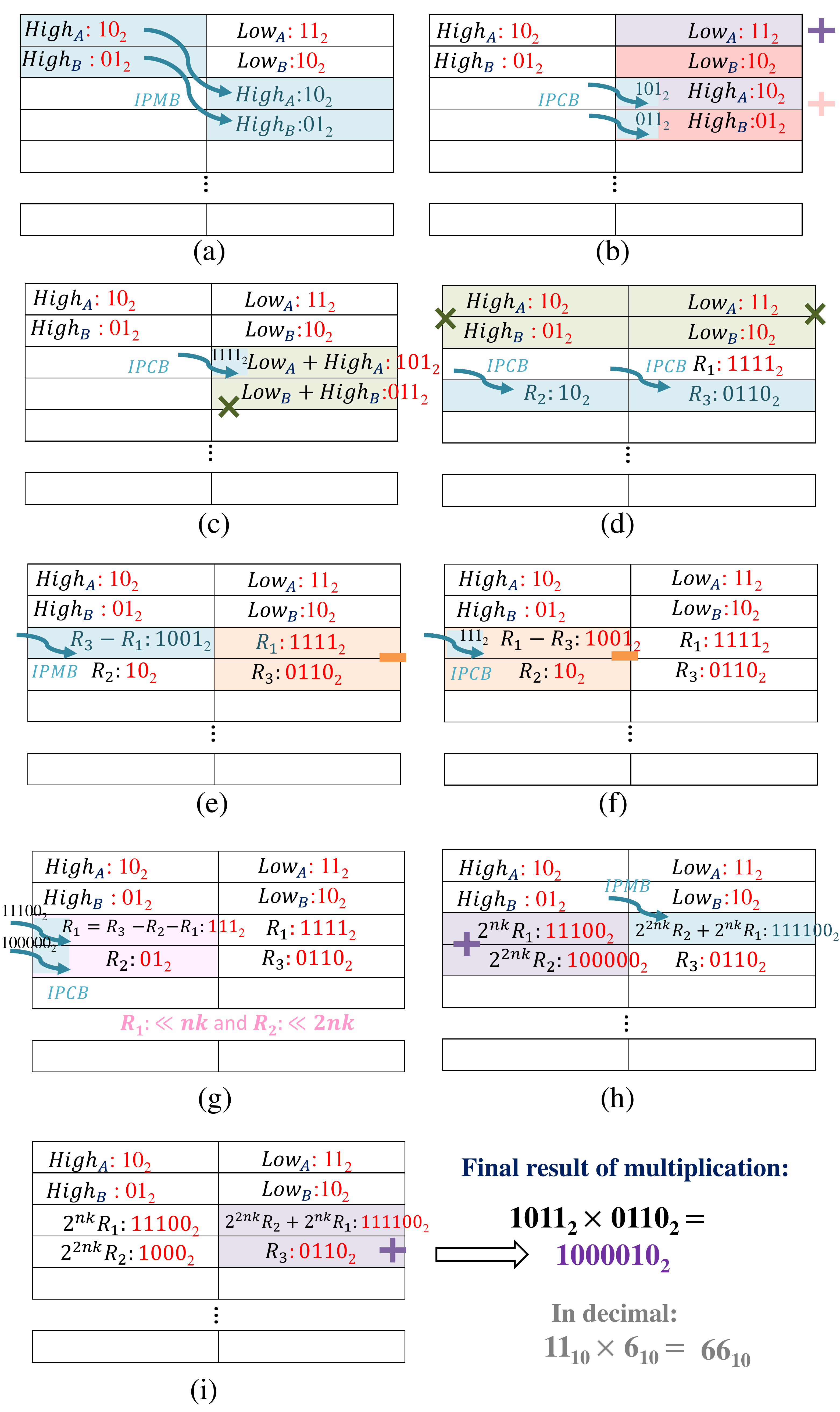}
    \caption{Mapping the Karatsuba multiplication for execution in memory. A toy example for multiplication of $A=11_{10}$ and $B=6_{10}$ ($A=1011_{2}$ and $B=0110_{2}$) is presented. Blue arrows (and shades) indicate the storage of intermediate computation results by IPCBs and IPMBs. Different colors, i.e., purple/red, green, orange and pink, are used to represent additions, coefficient-wise multiplications, subtractions and shifts, respectively. The flow of the algorithm execution is presented in eight parts. (a) Alignment of high and low parts of coefficients $A$ and $B$. (b) Computation of additions $Low_{A}+High_{A}$ and $Low_{B}+High_{B}$. (c) Computation of product $R1$. (d) Computation of products $R2$ and $R3$, which are performed in parallel. (e) Subtracting $R3$ from $R1$. (f) Subtracting $R2$ from $R1-R3$. (g) Left-Shifting $R1=R1-R2-R3$ by $nk$ and $R2$ by $2nk$. (h) Adding up the shifts results. (i) Computation of final addition $R2\times2^{2nk}+R1\times2^{nk}+R3$. }
    \label{fig:karatsuba_illustrated}
    \vspace{-3pt}
\end{figure}

The mapping of Karatsuba multiplication operations for execution \textit{in memory} are described with a toy example in Fig. \ref{fig:karatsuba_illustrated}. In our example, we want to multiply $A=11_{10}$ and $B=6_{10}$ ($A=1011_{2}$ and $B=0110_{2}$). In the first step of the algorithm, we first move the high part of both operands ($High_{A}=10_{2}$ and $High_{B}=01_{2}$) so that they are aligned to ($Low_{A}=11_{2}$ and $Low_{B}=10_{2}$), i.e., the high and low parts of the coefficients would map to the same columns, respectively. This step is illustrated in Fig. \ref{fig:karatsuba_illustrated}(a). We use IPMBs (described in Sec. \ref{sec:buffers} to move data from column $i$ to column $i+F$ in an array.

The second step consists of computing the additions $Low_{A}+High_{A}=11_{2}+10_{2}$ and $Low_{B}+High_{B}=10_{2}+01_{2}$ (Fig. \ref{fig:karatsuba_illustrated}(b)) with an in-memory adder (described in Sec. \ref{sec:CiM_hardware_description}). The results of the additions, i.e., $Low_{A}+High_{A}=101_{2}$ and $Low_{B}+High_{B}=011_{2}$, are copied to scratchpad addresses in memory with IPCBs (represented in Fig. \ref{fig:cim_hardware}(c)).

Next, we compute the product $R1$ between $Low_{A}+High_{A}=101_{2}$ and $Low_{B}+High_{B}=011_{2}$ using the \textit{Shift-Add} operation (Fig. \ref{fig:karatsuba_illustrated}(c)). The product $R1=1111_{2}$ is copied to the scratchpad space in memory with IPCBs. Similarly, we compute the other two products $R2=10_{2}$ and $R3=0110_{2}$ --- between low (and high) parts of $A$ and $B$ --- using the same procedure as for computing $R1$ (Fig. \ref{fig:karatsuba_illustrated}(d)).

The fifth and sixth steps of the in-memory Karatsuba multiplication (in Figs. \ref{fig:karatsuba_illustrated}(e) and \ref{fig:karatsuba_illustrated}(f)) are for computing the subtraction $R1-R2-R3=0111_{2}$. The subtraction is performed in two parts: first, we compute the two-term subtraction $R1-R3=1001_{2}$ using the in-memory subtraction as described in Sec. \ref{sec:poly_subtraction}. We use IPMBs to move the result of this subtraction so that it is aligned with product $R2=10_{2}$ (Fig. \ref{fig:karatsuba_illustrated}(e)). Next, we perform another subtraction of two-terms, i.e. between $R1-R3=1001_{2}$ and $R2=10_{2}$. The result of the subtraction $R1-R2-R3$, which we rename as $R1$, is copied to memory with IPCBs (Fig. \ref{fig:karatsuba_illustrated}(f)).

After we compute the subtraction, we shift: (i) $R1=0111_{2}$ by $nk$ bits, and (ii) $R2=10_{2}$ by $2nk$ bits using the log shifter (Fig. \ref{fig:karatsuba_illustrated}(g)). In this example, $k=2$ and $n=1$, i.e., the smaller product to be computed by the algorithm has 2-bit terms (base case). The results of the shifts are copied to scratchpad addresses in memory with IPCBs. Next, we perform an addition between $R1\times2^{nk}=11100_{2}$ and $R2\times2^{2nk}=100000_{2}$, and move the result ($111100_{2}$) so that it is aligned to $R3=0110_{2}$. The steps described are illustrated in Fig. \ref{fig:karatsuba_illustrated}(h). Last, we perform the final addition ($R2\times2^{2nk}+R2\times2^{nk}+R3=1000010_{2}$) (Fig. \ref{fig:karatsuba_illustrated}(i)). This result is equivalent to the product $A\times B$.

The dominant part of the execution time for this procedure is the multiplications between coefficients, i.e., the computation of products $R1$, $R2$ and $R3$. The other operations are additions, subtractions and copies (movement of data either within the same or different columns in memory). Due to the parallelism in CiM, we can compute the products $R2$ and $R3$ in parallel. In this scenario, the complexity of the Karatsuba algorithm can be reduced from $\mathcal{O}(n^{\log_2 3}) $ to $\mathcal{O}(n\log_2 n) $ \cite{cesari1996performance}.

Finally, note that the support of polynomial primitives that form the basis the execution of \textit{HomAdd}, \textit{HomSub}, and \textit{HomMult} operations enables CiM-HE to perform all computations required by HE.



\section{Evaluation}
\label{sec:evaluation}


We evaluate our CiM-HE architecture and compare the runtime and energy consumption of the three operations (\textit{HomAdd}, \textit{HomSub} and \textit{HomMult} in Sec. \ref{sec:CiM-HE_framework}) with a CPU-based HE implementation based on SEAL \cite{chen17_SEAL}. Furthermore, we evaluate three end-to-end tasks (arithmetic mean, variance, and linear regression) performed with both CiM-HE and the CPU. Finally, we compare the runtime and energy of \textit{HomMult} (the most expensive HE operation) with a state-of-the-art FPGA-based accelerator \cite{roy19_hpca}.

\subsection{Experimental setup}
\label{sec:evaluation_exp_setup}


The operations in the B/FV scheme are computed over a polynomial ring $\mathcal{R}_q$. 
We take $q$ to be a 218-bit number, (i.e. $\lceil \log_2 q \rceil=218$) and the degree of ciphertext polynomials to be $n=2^{13}\!=\!8192$. We choose these parameters to match the 128-bit security, one of the default settings of Microsoft's SEAL \cite{chen17_SEAL}, which exceeds the minimum security requirement of NIST \cite{barker2018transitioning} which is 112 bits. The security afforded by these parameters can be shown to be 128 bits with the estimator \cite{albrecht2015concrete}.  Consequently, each ciphertext contains two polynomials of degree $8192$ with 218-bit coefficients, and the ciphertext size is 436 KB. Furthermore,
 the multiplicative depth of the supported functions is $\!L=\!5$, which can be shown by the following inequality \cite{fan12_BFV}:
\begin{displaymath}
L < \frac{ log(\frac{\floor{\frac{q}{B}}}{4}) + log(t) - log(\delta_R + 1.25)}{log(\delta_R) + log(\delta_R+1.25) + log(t)}
\end{displaymath}
In our setting, the bound of the error distribution $B$ is 1, the expansion factor $\delta_R$ is $n$, and the plaintext modulus $t$ is $2^{10}$.



\subsubsection{CPU-based HE}
\label{sec:evaluation_exp_setup_CPU_HE}
To evaluate HE primitives on a CPU, we run \textit{HomAdd}, \textit{HomSub}, and \textit{HomMult} using the Microsoft SEAL library \cite{chen17_SEAL}. 
CPU-based HE performs best when the modulus $q$ is a product of distinct coprime numbers, as this allows residue number system (RNS) decomposition of coefficients as in \cite{bajard2016full} or \cite{halevi2019improved}. Because $2^k$ cannot be factored as a product of coprimes, using $2^k$ as a modulus does not allow for RNS decomposition of ciphertext coefficients. Thus using $2^k$ as a modulus would slow down computation on non-CiM systems.
SEAL thus automatically chooses $q$ to be a 218-bit number that is a product of 5 coprime numbers of 43 or 44 bits.
SEAL uses several algorithmic optimizations for HE, including coefficient-wise RNS decomposition as in \cite{bajard2016full}, and the Number-Theoretic Transform (NTT) for fast polynomial multiplication. 
In our CiM-HE system, we have not implemented any of these optimizations.
Thus, our basis for comparison for CiM-HE is the best case for the CPU. Despite the differences in the choice of the parameter $q$, (CiM-HE uses moduli in the form $2^k$, and SEAL uses a product of coprimes), the security level of both implementations is the same at 128 bits.

The configuration of our test machine is listed in Table \ref{tab:laptop_specs}. To account for runtime, we use the C++ std::chrono library. We use the Powerstat tool \cite{powerstat} to measure the power of the CPU while running the HE primitives. 
To offset the power consumption of system tasks not related to HE, we measure the idle power for the same amount of time as the execution of HE primitives and subtract it from the total power consumption. Finally, we calculate energy consumption as the product between HE net power and operation runtime.

\begin{table}[t]
\centering
\setlength\extrarowheight{2pt}
\caption{Specifications of CPU-based HE}
\begin{tabular}{|r|l|}
\hline
CPU Model: & Intel(R) Core(TM) i5-5300U CPU @ 2.30GHz \\ \hline
L1 cache: & 32 KB (L1i), 32KB (L1d) \\ \hline
L2 cache: & 256 KB \\ \hline
L3 cache: & 3 MB \\ \hline
RAM: & 8 GB DDR3 \\ \hline
\end{tabular}

\label{tab:laptop_specs}
\end{table}

\begin{figure}[!t]
    \centering
    \includegraphics[width=1\columnwidth]{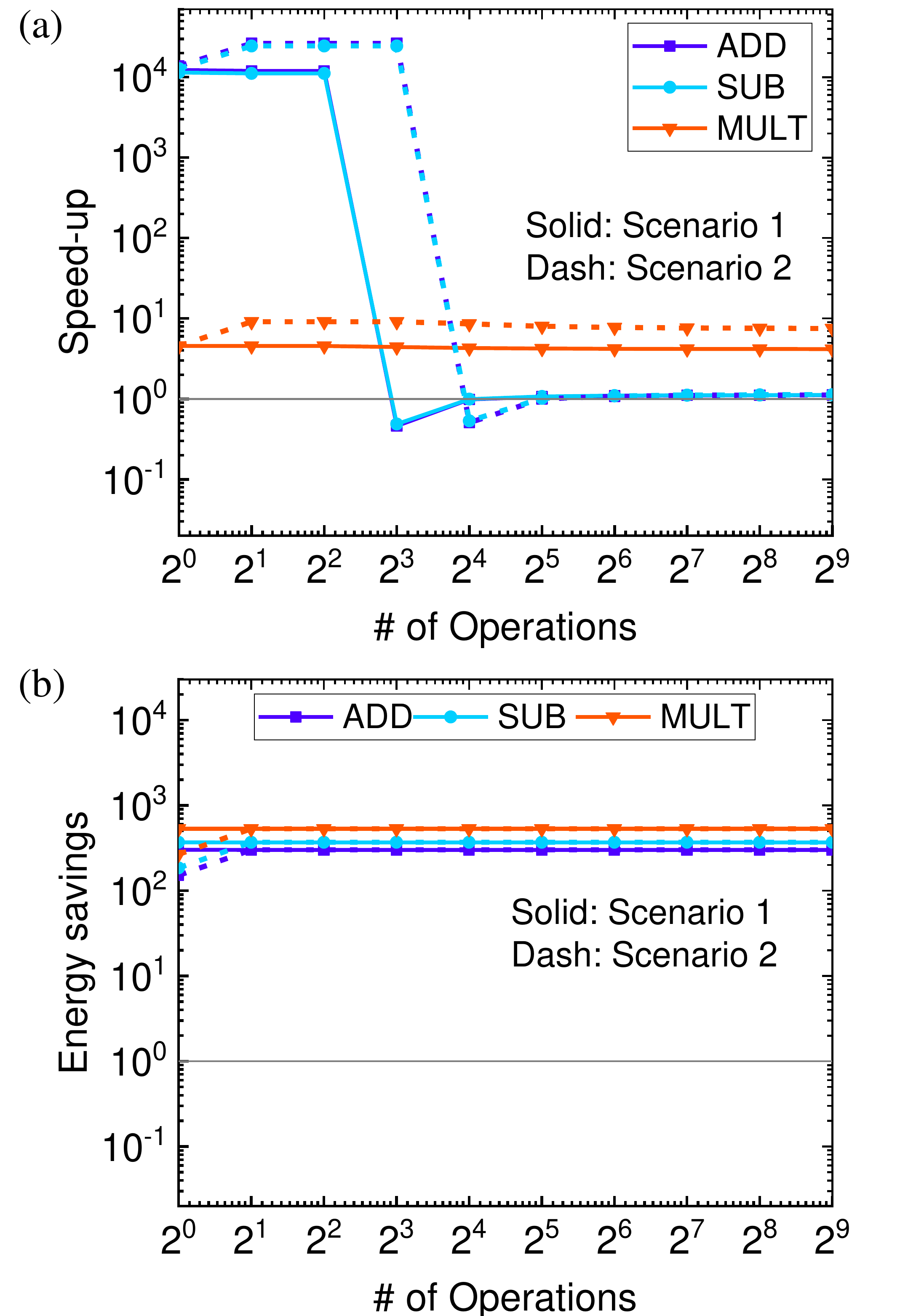}
    \caption{CiM-HE (a) speed-ups and (b) energy savings with respect to CPU. Solid (dash) lines represent scenarios 1 (and 2). For Scenario 2, CiM-HE banks perform operations in parallel increasing the throughput of CiM operations.}
    \label{fig:speedup_vs_ciphertexts}
\end{figure}


\subsubsection{CiM-HE}
\label{sec:evaluation_exp_setup_CiM_HE}
The time and energy of HE primitives implemented in CiM-HE are evaluated based on simulations of the architecture described in Sec. \ref{sec:CiM-HE_framework}. The modulus $q$ is set as a power of two, i.e. $q=2^{218}$, which is CiM-friendly. 
Memory arrays and peripherals are simulated in HSPICE \cite{hspice2018inc} using the 14nm BSIM-CMG FinFET model \cite{duarte15_bsim_cmg} (the same technology node as the CPU). The CiM controller and decoders are synthesized in Verilog with the Open Cell Library \cite{martins15_stdcell} at the same FinFET technology node. In circuit-level simulations, we measure the time and energy of executing each CiM step required by the polynomial operations (as described in Sec. \ref{sec:hardware_mapping}). Then, we determine the complete sequence of steps needed for each polynomial primitive. Based on this sequence, we compute the time and energy of all the polynomial primitives, which will ultimately be used to obtain the time and energy of each HE operation running on CiM-HE. 

\bluHL{The choice of where to place the CiM module is an important research topic that has been studied in previous works, e.g. \cite{aga17,gao2020eva}. For instance, one key finding of these studies is that placing CiM units at different levels in the memory hierarchy may result in distinct effects depending on the application. As CiM-HE performs computation at the memory periphery without the need for data transfers to external processing units, a key requirement for reasonable performance of the CiM-HE is that its size has to be sufficient to hold at least a pair of ciphertexts (operands), while leaving a “scratch space” to store intermediate results from the CiM operations.}

\bluHL{For the reason above, }our CiM architecture is implemented at the L3 cache level, which consists of 1 bank with 4096 arrays of size $8\times1024$ 6T-SRAM cells. Each array holds up to 32 coefficients of 218 bits, which can fit in 4 standard-sized words of 64 bits. 0.25 KB per array are reserved as scratch space to store intermediate results of CiM-HE operations. The total size of a CiM-HE bank is 4MB. To minimize the number of cycles needed for each CiM-HE primitive, we assume that coefficients of the same degree (of different ciphertexts) are stored in a column-aligned fashion (see Fig. \ref{fig:cim_hardware}(b) in Sec. \ref{sec:CiM_hardware_description} for more details). With the column-aligned placement scheme described in Sec. \ref{sec:CiM_hardware_description}, as well as the 4096 arrays in our CiM architecture, we are able to perform 2 simultaneous \textit{PolyAdd} operations, i.e., 8192 coefficient-wise additions in parallel for our HE parameter settings. \bluHL{Importantly, by implementing a CiM-HE of 4MB size (with 1MB reserved for scratch space) at the L3 cache, we are able to make a fair comparison of our work with the CPU baseline outlined in Table \ref{tab:laptop_specs}.}




\subsection{Single Execution of HE Operations}
\label{sec:evaluation_primitives}



We evaluate the single execution of HE operations (\textit{HomAdd}, \textit{HomSub} and \textit{HomMult}) on CiM-HE using the polynomial primitives described in Sec. \ref{sec:CiM-HE_framework}. In Table \ref{tab:single_exec}, we summarize the energy and runtime of these operations for CiM-HE and CPU-based implementations. We observe high speedups (up to 10426.5x) for operations that require only the execution of polynomial primitives between coefficients of the same degree, i.e. \textit{HomAdd} and \textit{HomSub}, as these operations can all be executed in parallel by CiM-HE components described in Sec. \ref{sec:CiM_hardware_description}.

A smaller speedup is observed for \textit{HomMult}, since this operation requires many steps to execute the required number of polynomial multiplications (\textit{PolyMult}). Note that the runtime of \textit{PolyMult} is proportional to both $log(q)$ and $n$, and this polynomial primitive requires {\bf (i)} multiplication between coefficients of the same degree (implemented with \textit{Shift-Add} operations), and {\bf (ii)} multiplications, additions and subtractions between coefficients of different ($n$) degrees. While CiM-HE is able to perform many coefficient-wise operations in parallel because of the inherently high internal bandwidth of the memory, the \textit{PolyMult} algorithm implemented on CiM relies on Karatsuba multiplication and does not employ more advanced optimizations such as NTT or RNS.

CiM-HE enables energy savings of 296.9x for \textit{HomAdd} and 532.8x for \textit{HomMult}. Besides in-memory additions, multiplications involve numerous shifts and copies, which consumes much less energy than additions when performed in memory. As such, shifts and copies lower the average power of \textit{HomMult} in comparison to \textit{HomAdd}. The high energy savings for multiplication with CiM-HE reflects its lower average power.


\begin{table}[t]
\centering
\setlength\extrarowheight{2pt}
\caption{Time and Energy of Single Execution of HE Operations}
\scalebox{.97}{\begin{tabular}{|l|c|c|c|c|c|c|}
\hline
\multicolumn{1}{|c|}{\multirow{2}{*}{Primitive}} & \multicolumn{3}{c|}{Time (s)} & \multicolumn{3}{c|}{Energy (J)} \\ \cline{2-7} 
\multicolumn{1}{|c|}{} & CPU & CiM & Imp. & CPU & CiM & Imp. \\ \hline
\textit{HomAdd} & 8.2E-5 & 7.9E-9 & \textbf{10426.5x} & 1.1E-3 & 3.6E-6 & \textbf{296.9x} \\ \hline
\textit{HomSub} & 8.7E-5 & 8.9E-9 & \textbf{9790.6x} & 1.4E-3 & 3.9E-6 & \textbf{364.5x} \\ \hline
\textit{HomMult} & 3.0E-2 & 6.6E-3 & \textbf{4.6x} & 3.7E-1 & 7.0E-4 & \textbf{532.8x} \\ \hline
\end{tabular}}
\label{tab:single_exec}
\end{table}


\begin{figure*}[!t]
    \centering
    \includegraphics[width=2\columnwidth]{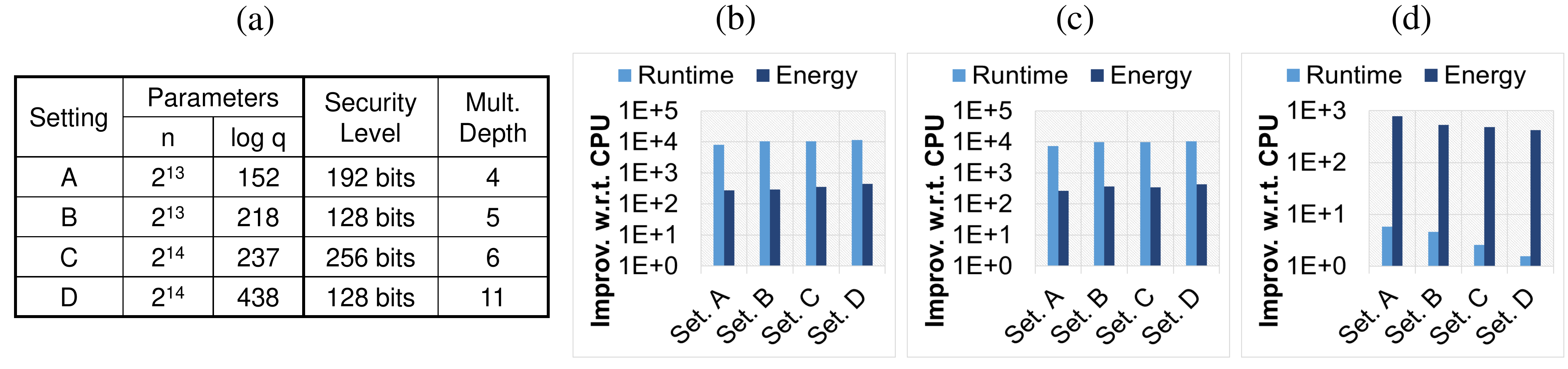}
    \caption{Our evaluation of \textbf{(a)} four different parameter settings show the runtime and energy improvements of homomorphic \textbf{(b)} additions, \textbf{(c)} subtractions, and \textbf{(d)} multiplications when compared to a CPU implementation.}
    \label{fig:parameter_sweep}
\end{figure*}

\subsection{Multiple Executions of HE Operations}
\label{sec:evaluation_multiple_execution}
Here, we consider multiple executions of HE operations for different numbers of CiM-HE banks and different sizes of a CPU's L3 cache to evaluate the impact of data transfers on the runtime and energy efficiency of HE. Our evaluation is divided into two scenarios as described below:

\begin{itemize}
    \item \textit{Scenario 1}: We consider a single CiM-HE bank of size 4MB, and further assume a CPU's L3 cache of 4MB.\footnote{We estimate 4MB SRAM access time and energy with NVSim \cite{dong12_nvsim}. Cache hit, miss, and write latencies (energies) are 1.189ns (0.949nJ), 0.286ns (0.949nJ), and 0.621ns (0.903nJ) per access, respectively.} We operate on $N$ ciphertexts. Only 6 (8) ciphertexts are present in CiM-HE (CPU's L3 cache) initially. The rest $N\!-\!6$  ($N\!-\!8$) ciphertexts must be fetched from DRAM.\footnote{We consider a DRAM access time of 100ns and 80pJ/bit energy consumption \cite{borkar2010exascale}. Data blocks are 64B. }
    \item \textit{Scenario 2}: We consider two CiM-HE banks of size 4MB (a total size of 8MB), and further assume a CPU's L3 cache of 8MB.\footnote{We estimate 8MB SRAM access time and energy with NVSim \cite{dong12_nvsim}. Cache hit, miss, and write latencies (energies) are 2.075ns (1.230nJ), 0.339ns (1.230nJ), and 1.173ns (1.156nJ) per access, respectively.} We operate on $N$ ciphertexts. Only 12 (16) ciphertexts are present in CiM-HE (CPU's L3 cache) initially. The rest $N\!-\!12$  ($N\!-\!16$) ciphertexts must be fetched from DRAM.
\end{itemize} 

    

These two different scenarios are evaluated with an increasing number of HE operations on ciphertexts, i.e. $2^0$---$2^{9}$. \textit{CiM-HE} speedup (and energy savings) for Scenarios 1 and 2 are depicted in Fig. \ref{fig:speedup_vs_ciphertexts}(a) (and Fig. \ref{fig:speedup_vs_ciphertexts}(b)). The CiM-HE-based L3 cache is filled with 6 ciphertexts on which we initially operate (i.e., the cache is at its full capacity, considering that scratch space is needed to store intermediate results from the execution of HE algorithms). Likewise, the CPU's L3 cache of 4MB (iso-capacity) stores 8 ciphertexts as no scratch space is needed. We define the speedup as $({T_{CPU}+MT_{CPU}})/({T_{CiM}+MT_{CiM}})$, where $T_{CPU}$ ($T_{CiM}$) refers to execution time on CPU (CiM) and $MT_{CPU}$ ($MT_{CiM}$) refers to the time required to transfer data from DRAM to the L3 cache/CiM. 

With a single CiM-HE bank of 4MB and equivalent size of CPU's L3 cache (Scenario 1), the maximum speedups (and energy savings) depicted in Fig. \ref{fig:speedup_vs_ciphertexts}(a) (Fig. \ref{fig:speedup_vs_ciphertexts}(b)) are 12308.2x (301.2x), 11449.9X (368.4X) and 4.6x (532.8x) for \textit{Add}, \textit{Sub}, and \textit{Mult}, respectively. When the number of operations is larger than 6 for CiM-HE (and 8 for CPU), more ciphertexts need to be fetched from DRAM to replace the ciphertexts already present in the cache. For instance, when we perform 8 operations, CiM needs to fetch 2 ciphertexts from main memory, while the CPU's L3 cache already contains all the ciphertexts needed for computation. The difference in the number of ciphertexts present in the two L3 caches causes the CiM-HE speedup to drop below 1 for all operations (except \textit{Mult}) when the number of operations is equal to 8. For a larger number of operations, the speedup of \textit{HomAdd} and \textit{HomSub} stays close to 1, because $MT_{CPU}\gg T_{CPU}$, and $MT_{CiM}\gg T_{CiM}$ for these primitives. \textit{HomMult} can sustain its speedups at the same level because its computation time is more significant than the time spent on data transfers, i.e., for \textit{HomMult}, $T_{CPU}\gg MT_{CPU}$ and $T_{CPU}\gg MT_{CPU}$. Unlike speedup, the energy savings of CiM-HE for all operations (depicted in Fig. \ref{fig:speedup_vs_ciphertexts}(b)) do not suffer a significant drop when there are data transfers from DRAM to cache. This is because the energy spent on data transfers is much smaller than the energy spent on HE computations. 

With an 8MB L3 cache (Scenario 2), the CPU stores 16 ciphertexts (twice as many as in Scenario 1), while CiM-HE stores 12 ciphertexts (6 in each CiM-HE bank of 4MB). Therefore, the CPU has an advantage of 4 more ciphertexts that are cached before the need for fetching from DRAM. By using 2 CiM-HE banks, two HE operations can be performed in parallel (a similar scenario as in \cite{roy19_hpca}), which leads to further improvements with CiM-HE when compared to the CPU (dash lines in Fig. \ref{fig:speedup_vs_ciphertexts}(a)). Namely, we achieved maximum speedups (and energy savings) of 26386.7x (301.1X), 24461.1x (368.3X) and 9.1x (532.8x) for \textit{Add}, \textit{Sub} and \textit{Mult}, respectively.

As in Scenario 1, in Scenario 2 we observe a drop in speedup (but not in energy savings) when we start fetching data from DRAM, as the runtime is dominated by data transfers. Using more than one CiM-HE bank (Scenario 2) \textit{does not} improve the runtime or energy of each HE operation, and may not be advantageous if the system always performs one HE operation at a time (albeit unlikely in real applications). However, as an extra CiM-HE bank nearly doubles the throughput of the system because of the higher level of parallelism, the speedup for multiple HE operations is also improved. For instance, the maximum throughput for \textit{Mult} (the most expensive HE operation) rises from 151 multiplications per second with 1 CiM-HE bank to 302 multiplications per second with 2 CiM-HE banks working in parallel.  The energy consumption for 2 CiM-HE banks (Scenario 2) remains at same level as with a single CiM-HE bank (Scenario 1) when we perform multiple HE operations.

Note that the runtimes of \textit{HomMult} are much higher than other primitives, and the overall runtimes are dominated by \textit{PolyMult}. In complex computation tasks with various operations, we expect to have a meaningful speedup even if we have to bring a large number of ciphertexts into the cache, and will show this in the next subsection.

\subsection{Different parameter settings in CiM-HE}
\label{sec:parameter_sweeping}

\bluHL{As demonstrated in Fig. \ref{fig:cim_hardware}(a) and Fig. \ref{fig:cim_hardware}(b), CiM-HE can support various parameter settings for the B/FV encryption scheme by leveraging the appropriate ciphertext-to-memory mapping. Different parameter settings result in different security levels and multiplicative depths. A higher multiplicative depth (e.g., Setting D) enables the evaluation of a larger number of nested multiplications without the need for decryption or bootstrapping. Higher multiplicative depths can be useful for applications such as training/inference with neural networks of medium/high complexity We carry out an evaluation for homomorphic operations (\textit{HomAdd}, \textit{HomSub}, and \textit{HomMult}) with four different parameter settings (A through D), as listed in Fig. \ref{fig:parameter_sweep}(a). In this evaluation, we employ a CiM size of 8 MB (i.e., 8192 CiM-HE arrays). }

\bluHL{Figs. \ref{fig:parameter_sweep}(b-d) depict the runtime and energy improvements for a single execution of \textit{HomAdd}, \textit{HomSub}, and \textit{HomMult}, respectively, when compared to a CPU-based implementation with the same parameters. Note that the mapping of the ciphertexts to CiM-HE needs to ensure that polynomials' coefficients are placed in a column-aligned fashion to enable in-memory computing at the bitline level. This requirement determines the amount of parallelism possible with CiM-HE when computing on multiple polynomials. Operations with large ciphertexts (a result of higher security levels or multiplicative depths) tend to be more disadvantaged by the mapping requirement due to less parallelism. For instance, while Setting A enables 6 simultaneous polynomial operations with CiM-HE, Setting D only admits 1 polynomial operation at a time.}

\bluHL{The aforementioned mapping requirements do not affect the runtime and energy improvements for all homomorphic operations equally. First, as polynomial additions and subtractions (plus modular reduction) are a direct implementation of \textit{HomAdd} and \textit{HomSub}, operating on ciphertexts of different sizes does not substantially affect the runtime and energy efficiency of CiM-HE for these operations. This observation holds even when there is a need for serializing \textit{HomAdd} and \textit{HomSub} operations through 2 rounds of \textit{PolyAdd} and \textit{PolySub} (Setting D). However, improvements for the \textit{HomMult} operation are more susceptible to differences in the parameter setting, as a homomorphic multiplication cannot be directly implemented by a single \textit{PolyMult}. Conversely, \textit{HomMult} requires multiple rounds of \textit{PolyMult} along with other operations to be performed. Hence, it is more sensitive to the serialization of polynomial primitives.}

\subsection{Area evaluation}
\label{sec:area_evaluation}

\begin{figure}[!t]
    \centering
    \includegraphics[width=1\columnwidth]{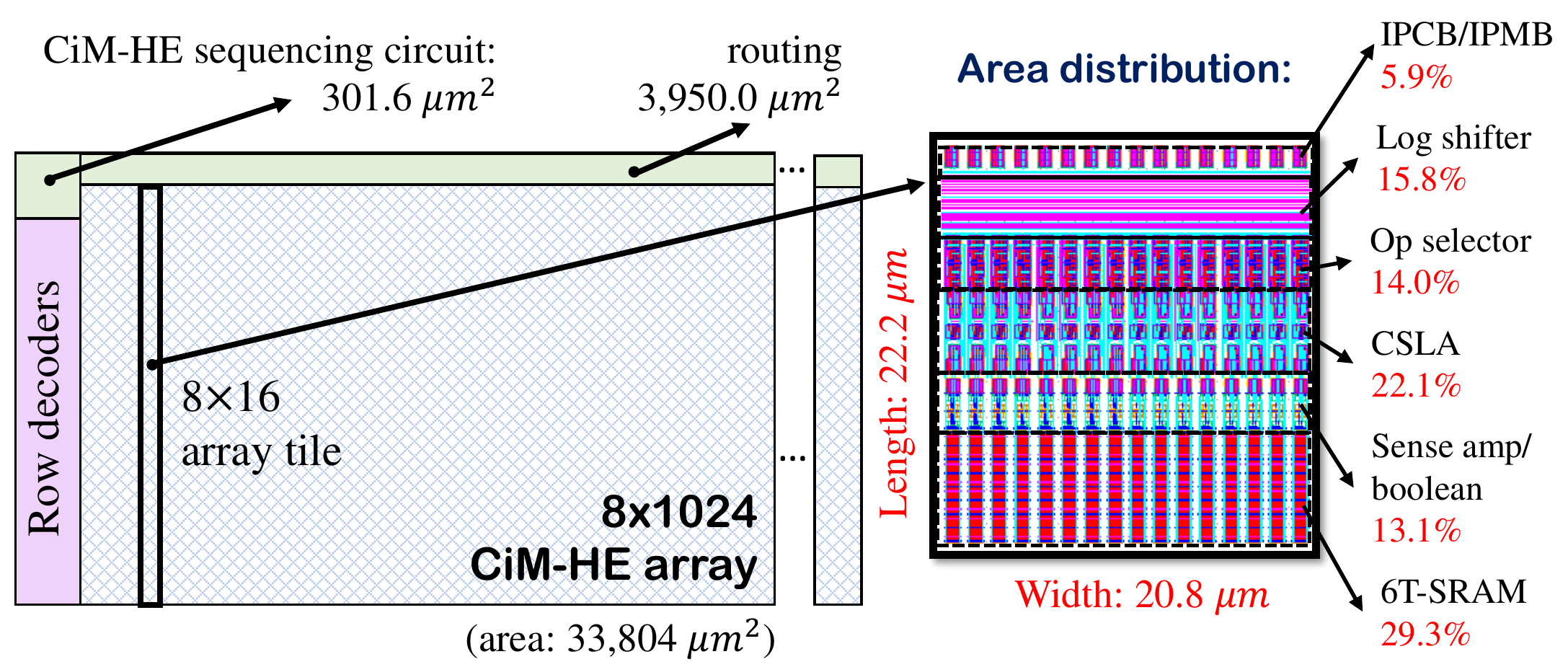}
    \caption{Area evaluation of a CiM-HE array.}
    \label{fig:area_eval}
\end{figure}

We estimate the area of one CiM-HE array of 8$\times$1024 size based on the modular layout of a 8$\times$16 tile, as depicted in Fig. \ref{fig:area_eval}. We use Cadence Virtuoso with FreePDK15 design kit \cite{bhanushali2015freepdk15} to construct the layout. Furthermore, we estimate the area of the sequencing circuit from the synthesized Verilog netlist with Cadence Encounter. The area of row decoders is not included in our evaluation, as they are standard elements in a memory and not exclusive to CiM-HE. Based on the modular design, the area of 1 CiM-HE array corresponds to 64$\times$ the area of one 8$\times$16 tile. When the sequencing circuits (and their respective routing overhead) are included, the resulting area is 33,804 $\mu m^{2}$. Per Fig. \ref{fig:area_eval}, the CiM components of CiM-HE occupy 70.7\% of the array tile area. 




\begin{table} [!t]
\setlength\extrarowheight{2pt}
\caption{Evaluation of \textit{Mean}, \textit{Variance}, and \textit{LinReg} tasks over (\subref{tab:end_to_end_ideal_locality}) 6 and (\subref{tab:end_to_end_bad_locality}) 60 ciphertexts for Scenario 1. Number of dimensions for \textit{LinReg} is 4.}
\begin{subtable}{1\columnwidth}
\centering

\scalebox{.97}{\begin{tabular}{|l|c|c|c|c|c|c|}
\hline
\multicolumn{1}{|c|}{\multirow{2}{*}{Task}} & \multicolumn{3}{c|}{Time (s)} & \multicolumn{3}{c|}{Energy (J)} \\ \cline{2-7} 
\multicolumn{1}{|c|}{} & CPU & CiM & Imp. & CPU & CiM & Imp. \\ \hline
Mean & 4.7E-4 & 3.9E-8 & \textbf{11985.4x} & 5.4E-3 & 1.8E-5 & \textbf{300.1x} \\ \hline
Variance & 1.5E-1 & 3.3E-2 & \textbf{4.6x} & 2.1E+0 & 5.3E-3 & \textbf{404.9x} \\ \hline
LinReg & 4.5E+0 & 9.9E-1 & \textbf{4.6x} & 5.6E+1 & 1.1E-1 & \textbf{532.3x} \\ \hline
\end{tabular}}
\caption{Scenario 1, file size of 6 ciphertexts}
\label{tab:end_to_end_ideal_locality}
\end{subtable}
\par

\begin{subtable}{1\columnwidth}
\centering
\scalebox{1}{\begin{tabular}{|l|c|c|c|c|c|c|}
\hline
\multicolumn{1}{|c|}{\multirow{2}{*}{Task}} & \multicolumn{3}{c|}{Time (s)} & \multicolumn{3}{c|}{Energy (J)} \\ \cline{2-7} 
\multicolumn{1}{|c|}{} & CPU & CiM & Imp. & CPU & CiM & Imp. \\ \hline
Mean & 4.7E-2 & 4.3E-2 & \textbf{1.1x} & 6.4E-2 & 2.1E-4 & \textbf{301.1x} \\ \hline
Variance & 2.1E+0 & 4.8E-1 & \textbf{4.4x} & 2.6E+1 & 6.4E-2 & \textbf{404.6x} \\ \hline
LinReg & 6.7E+2 & 1.7E+2 & \textbf{4.1x} & 8.1E+3 & 1.5E+1 & \textbf{532.4x} \\ \hline
\end{tabular}}
\caption{Scenario 1, file size of 60 ciphertexts}
\label{tab:end_to_end_bad_locality}
\end{subtable}
\label{tab:end_to_end1}
\end{table}


\begin{table} [!t]
\setlength\extrarowheight{2pt}
\caption{Evaluation of \textit{Mean}, \textit{Variance}, and \textit{LinReg} tasks over (\subref{tab:end_to_end_ideal_locality}) 6 and (\subref{tab:end_to_end_bad_locality}) 60 ciphertexts for Scenario 2. Number of dimensions for \textit{LinReg} is 4.}
\begin{subtable}{1\columnwidth}
\centering

\scalebox{.97}{\begin{tabular}{|l|c|c|c|c|c|c|}
\hline
\multicolumn{1}{|c|}{\multirow{2}{*}{Task}} & \multicolumn{3}{c|}{Time (s)} & \multicolumn{3}{c|}{Energy (J)} \\ \cline{2-7} 
\multicolumn{1}{|c|}{} & CPU & CiM & Imp. & CPU & CiM & Imp. \\ \hline
Mean & 5.2E-4 & 2.0E-8 & \textbf{26386.7x} & 5.4E-3 & 1.8E-5 & \textbf{299.7x} \\ \hline
Variance & 1.5E-1 & 1.6E-2 & \textbf{9.2x} & 2.1E+0 & 5.3E-3 & \textbf{404.9x} \\ \hline
LinReg & 4.5E+0 & 4.9E-1 & \textbf{9.1x} & 5.6E+1 & 1.1E-1 & \textbf{532.3x} \\ \hline
\end{tabular}}
\caption{Scenario 2, file size of 6 ciphertexts}
\label{tab:end_to_end_ideal_locality}
\end{subtable}
\par

\begin{subtable}{1\columnwidth}
\centering
\scalebox{1}{\begin{tabular}{|l|c|c|c|c|c|c|}
\hline
\multicolumn{1}{|c|}{\multirow{2}{*}{Task}} & \multicolumn{3}{c|}{Time (s)} & \multicolumn{3}{c|}{Energy (J)} \\ \cline{2-7} 
\multicolumn{1}{|c|}{} & CPU & CiM & Imp. & CPU & CiM & Imp. \\ \hline
Mean & 4.1E-2 & 3.9E-2 & \textbf{1.1x} & 6.4E-2 & 2.1E-4 & \textbf{301.1x} \\ \hline
Variance & 2.1E+0 & 2.8E-1 & \textbf{7.7x} & 2.6E+1 & 6.4E-2 & \textbf{404.6x} \\ \hline
LinReg & 6.7E+2 & 9.5E+1 & \textbf{7.1x} & 8.1E+3 & 1.5E+1 & \textbf{532.4x} \\ \hline
\end{tabular}}
\caption{Scenario 2, file size of 60 ciphertexts}
\label{tab:end_to_end_bad_locality}
\end{subtable}
\label{tab:end_to_end2}
\end{table}

\subsection{End-to-End Tasks}
\label{sec:end_to_end_tasks}

One potential application of CiM-HE is in secure computation on private data. One use case of CiM-HE is that several parties jointly compute a function on their secret inputs (e.g., hospitals jointly computing statistics on patient data). Another use case is that a client employs a machine learning model in a trusted cloud computing service for inference without disclosing the private data. In these examples, the parties can encrypt their data, allow the cloud computing service to perform the computation homomorphically, then decrypt the results locally. To evaluate the use cases of CiM-HE in real-life situations, we homomorphically compute the functions described below.

    \subsubsection{\textbf{Arithmetic mean $\mu = \frac{1}{n}\sum_{i=1}^n x_i$}}
    \label{sec:arithmetic_mean}
      CiM-HE computes and returns the encrypted sum, which is decrypted and divided by the client.
    
    \subsubsection{\textbf{Variance  $\sigma^2 = \frac{1}{n}\sum_{i=1}^n (x_i-\mu)^2$}}
    \label{sec:variance}
    CiM-HE computes the value of $\sigma^2$ equivalently as $\sigma^2 = \frac{1}{n^3}\sum_{i=1}^n(n\cdot x_i-\sum_{j=1}^n x_j)^2$. The clients performs the final decryption and division by $n^3$.
    
    \subsubsection{\textbf{Arithmetic operations of linear regression}}
    \label{sec:linear_regression}
    Given a set of arguments $\mathbf{X} \in \mathbb{F}_{N\times D}$ and target values $\mathbf{t} \in \mathbb{F}_{1 \times D}$ over a field $\mathbb{F}$, the optimal weights for linear regression are calculated by $\mathbf{w} = (\mathbf{X^T X)^{-1}X^T t}$. We examine only the arithmetic operations required in calculating $\mathbf{w}$, and do not consider the matrix transpositions and inversions.
    
    \subsubsection{\textbf{Inference with a MLP neural network}}
    \label{sec:mlp_inference}
    
\begin{figure}[!t]
    \centering
    \includegraphics[width=1\columnwidth]{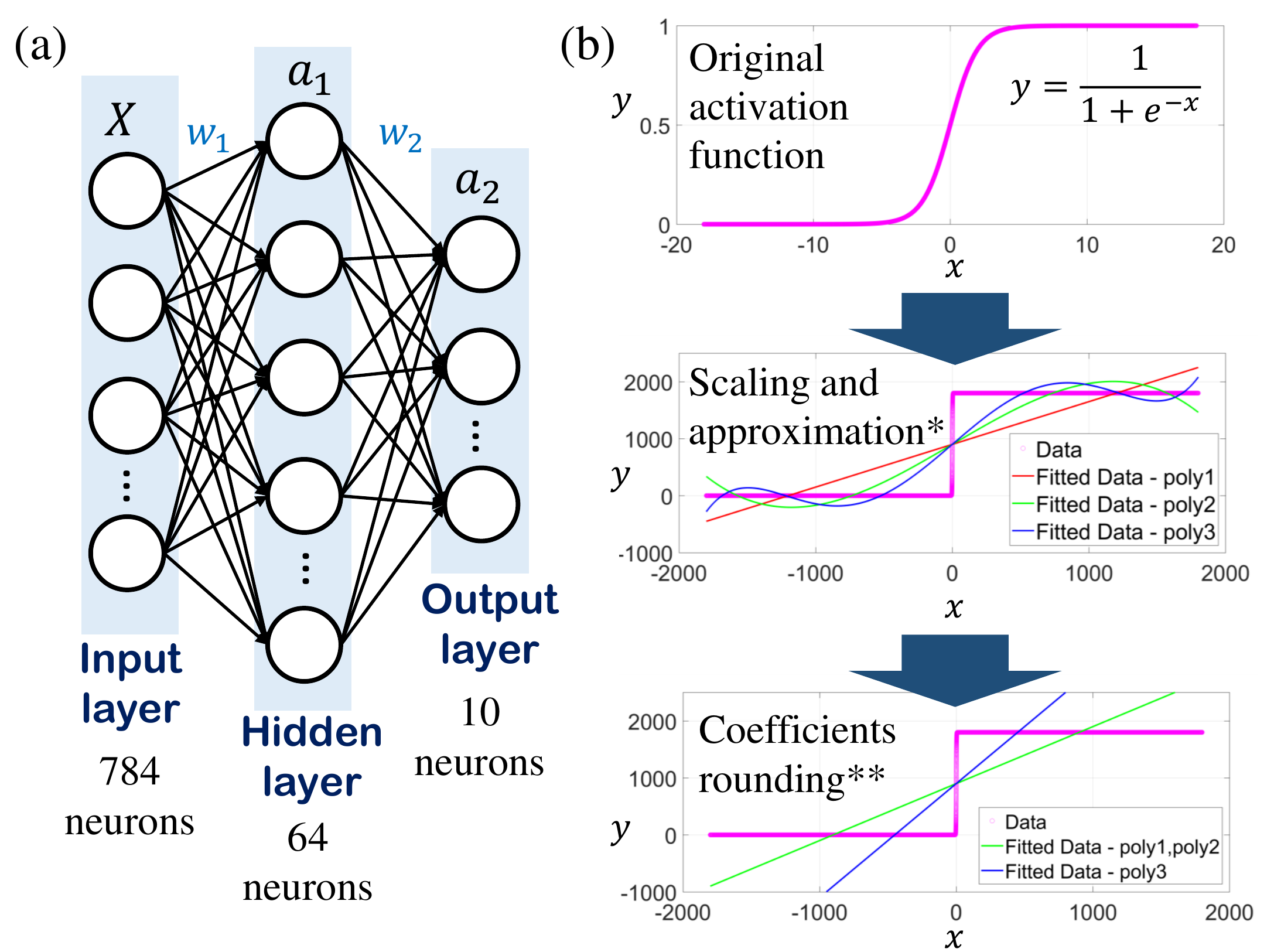}
\begin{flushright}
\scriptsize
\textbf{*}Three polynomial approximations (with degrees 2, 3, and 5) are found:\\
\textbf{poly1:} $-7.7 e^{-19}x^{2}+0.8x+900$\\
\textbf{poly2:} $-3.4 e^{-7}x^{3}+1.7 e^{-18}x^{2}+1.4x+900$\\
\textbf{poly3:} $2.6 e^{-13}x^{5}-2.5 e^{-24}x^{4}-1.3 e^{-6}x^{3}+6.1 e^{-18}x^{2}+2.1x+899$
\textbf{**}After the polynomial coefficients are rounded to the nearest integer, the approximations become:\\
\textbf{poly1,poly2:} $1x+900$\\
\textbf{poly3:} $2x+899$\\
\end{flushright}
    \caption{(a) Structure of the 3-layer MLP neural network. (b) Original sigmoid activation function and adjustments to enable inference on encrypted data with B/FV scheme.}
    \label{fig:mlp_network}
    \vspace{-2ex}
\end{figure}

\begin{table} [!t]
\caption{Inference evaluation}
\centering
\setlength\extrarowheight{2pt}
\begin{tabular}{|l|c|c|c|c|c|c|}
\hline
\multicolumn{1}{|c|}{\multirow{2}{*}{Cache size}} & \multicolumn{3}{c|}{Time (s)} & \multicolumn{3}{c|}{Energy (J)} \\ \cline{2-7} 
\multicolumn{1}{|c|}{} & CPU & CiM & Imp. & CPU & CiM & Imp. \\ \hline
4MB & 1.6E+3 & 3.8E+2 & \textbf{4.2x} & 1.9E+4 & 3.6E+1 & \textbf{532.8x} \\ \hline
8MB & 1.6E+3 & 2.1E+1 & \textbf{7.5x} & 1.9E+4 & 3.6E+1 & \textbf{532.8x} \\ \hline
\end{tabular}
\label{tab:inference}
\end{table}
    
    Given a 3-layer MLP (Fig. \ref{fig:mlp_network}(a)), CiM-HE performs inference on the encrypted MNIST data set of handwritten numbers \cite{deng2012mnist}. The MLP neural network was previously trained with unencrypted data. The sigmoid activation function, i.e., $Y = 1/(1+e^{-x})$, was employed for training. This baseline, 3-layer MLP neural network achieves 93.1\% accuracy at inference.\footnote{Higher baseline accuracies might be possible by fine adjusting the MLP network and its parameters, e.g., the learning rate, epochs, number of neurons in the hidden layer etc. However, in this study we primarily focus on establishing an \textbf{initial} baseline that is meant to guide future study at scale for SHE applied to machine learning problems.} The fact that the B/FV SHE scheme operates on polynomials with integer coefficients require the weights and the activation function to be adjusted for inference on encrypted data. The adjustments are listed below. Note that after we perform these adjustments, the 3-layer MLP neural network can perform inference on encrypted MNIST data set without the need for bootstrapping.
    
    \begin{itemize}
    \item \textbf{Adjustment 1: }A scaling factor\footnote{\label{first}Our evaluation assumes $F_{w}=100$ and $F_{a}=1800$, which yielded the best training accuracy among several different combinations tested.} $F_{w}$ was applied to the weights post training to ensure that their values are integers;
    \item \textbf{Adjustment 2: }A scaling factor\cref{first} $F_{a}$ was applied to the activation function, which becomes: $Y = F_{a}\times(1/(1+e^{-x}))$. The scaled sigmoid function is approximated by polynomial functions (Fig \ref{fig:mlp_network}(b)). The coefficients of the polynomials must be rounded to the nearest integer.  The linear function $y=1x+900$ is found to be a possible solution to the approximation, and it is used in our evaluation. The function yields 84\% testing accuracy versus 77\% when using another option available, i.e., $y=2x+899$.
    \end{itemize}

\subsubsection{\textbf{Result Discussion}}
    \label{sec:result_discussion}

The results of our evaluation for the \textit{Mean}, \textit{Variance}, and \textit{LinReg} tasks are presented in Tables \ref{tab:end_to_end1} and \ref{tab:end_to_end2}. Two different situations are considered regarding \textbf{file sizes}, which represent the number of ciphertexts (inputs) to be processed in each one of these tasks. Namely, we evaluate \textit{Mean}, \textit{Variance}, and \textit{LinReg} tasks with CiM-HE and CPU considering:
\begin{itemize}
    \item A file size of 6 ciphertexts (Table  \ref{tab:end_to_end1}(a) for Scenario 1 and Table \ref{tab:end_to_end2}(a) for Scenario 2)
        \item A file size of 60 ciphertexts (Table  \ref{tab:end_to_end1}(b) for Scenario 1 and Table \ref{tab:end_to_end2}(b) for Scenario 2)
\end{itemize}

The \textit{Inference} task has a fixed input size, i.e., an image with $28\times28$ pixels, therefore we present the results of our evaluation for this task in a separate table (Table \ref{tab:inference}). When encrypted, the input corresponds to 784 ciphertexts that encode 1 handwritten digit. Note that the 4MB and 8MB caches are not large enough to hold all the ciphertexts at inference time, so data transfers occur for the two different CiM size scenarios during inference. 


Computing the \textit{Mean}, \textit{Variance}, and \textit{LinReg} with a file size of 6 ciphertexts requires no data fetches in either Scenario 1 or Scenario 2, as all ciphertexts are present in the L3 caches per Sec. \ref{sec:evaluation_multiple_execution}. More than five orders of magnitude speedup ($>$11,000x with Scenario 1 and $>$26,000x with Scenario 2) are observed for the \textit{Mean} task as only \textit{HomAdd} operations are executed. The execution time of \textit{HomMult} dominates others in variance and linear regression tasks, i.e., the speedup with Scenario 1 for \textit{Variance} and \textit{LinReg} is 4.6x, which is analogous to the speedup of \textit{HomMult} alone.

On the other hand, executing \textit{Mean}, \textit{Variance}, and \textit{LinReg} for a file size of 60 ciphertexts causes the improvement for the \textit{Mean} task to drop to 1.1X in either scenario as the time to fetch ciphertexts dominates computational time. As expected, for other tasks where multiplication operations dominate runtime, we observed a minimum speedup of 4.4x and 4.1x for \textit{Variance} and \textit{LinReg} with Scenario 1.

The runtime of the \textit{Inference} task is also dominated by multiplications, which explains the 4.2x (7.5x) runtime improvements of CiM-HE when compared to CPU for cache sizes of 4MB (8MB). While doubling the memory size does not change the runtime of the CPU-based solution, it allows for more parallel computations to happen in two distinct CiM banks (similar to the two co-processors solution in \cite{roy19_hpca}).

Energy improvements reach two orders of magnitude for all the tasks evaluated, i.e., we obtain a minimum of $>$290x energy savings for \textit{Mean} and maximum of $>$500x energy savings for \textit{LinReg} and \textit{Inference}. CiM-HE consumes more energy in \textit{Mean} than in other tasks. Calculating a sum homomorphically with \textit{HomAdd} requires only additions between coefficients of the same degree, which can be performed with the fast in-memory carry select adders in our CiM-HE architecture. Other tasks, e.g., \textit{Variance}, \textit{LinReg}, and \textit{Inference} are dominated by multiplications (\textit{HomMult}). As shown in Sec. \ref{sec:evaluation_primitives}, \textit{HomAdd} has higher power consumption than \textit{HomMult}, hence the lower energy improvement for tasks that involve more additions.

\subsection{Comparison with Related Works}
\label{sec:comparison_with_FPGA}

Here, we compare the performance and energy efficiency of \textit{HomMult} running on CiM-HE with previous works that propose HE implementations \cite{bos2017privacy,roy19_hpca}. As highlighted in \cite{roy19_hpca}, homomorphic multiplications are the primary bottleneck of HE due to their extremely long runtime. As such, a comparison for this primitive alone is sufficient to assess the benefits of our proposed CiM-HE when compared to existing work. Reference \cite{bos2017privacy} is a CPU implementation based on the NFLlib \cite{cryptoexperts} that employs the B/FV scheme. Reference \cite{roy19_hpca} proposes an accelerator that employs up to 2 FPGA-based co-processors for HE, and implements NTT and RNS for homomorphic multiplications that are also based on the B/FV scheme.

An important challenge of making a fair comparison between CiM-HE and other works is that each HE implementation uses a different set of parameters, which can significantly impact security, multiplicative depth and runtime of HE operations. For this reason, the parameters $\log{q}$ and $n$, as well as the use of NTT and RNS optimizations in \cite{bos2017privacy,roy19_hpca} and this work are listed in Table \ref{tab:comparison_other_works}. The security level resulted from these parameters is only 80 bits, which has not been considered to be acceptable by NIST since 2015 \cite{barker2018transitioning}. Nevertheless, we use a similar parameter setting in our design so as to make a fair comparison between CiM-HE and the existing work \cite{bos2017privacy, roy19_hpca}. 

Table \ref{tab:comparison_other_works2} presents the homomorphic multiplication time and energy for \cite{bos2017privacy,roy19_hpca} and CiM-HE. Roy, et al. employ their faster configuration with 2 FPGA-based co-processors. Bos, et al. run their HE implementation on an Intel Core i5-3427 CPU at 1.8 GHz. Per \cite{intelcpu}, latest generation Intel i5 reaches up to 40W on heavy load operations, which we assume when comparing energy consumption of \cite{bos2017privacy} with our proposed CiM-HE. The same assumption was made by \cite{roy19_hpca}.

Execution of a homomorphic multiplication takes 5.0 ms and 33.0 ms in \cite{roy19_hpca} and \cite{bos2017privacy}, respectively. CiM-HE performs the same operation in 2.3 ms, which represents a speedup of 14.3x with energy savings of $>$2600x when compared to \cite{bos2017privacy}. The use of 2 FPGA-based co-processors (1 CiM-HE bank that processes two HE operations in parallel) allows for a throughput of 400 (861) multiplications per second in \cite{roy19_hpca} (CiM-HE). Furthermore, CiM-HE enables 88.1x more energy savings, with same the security level and multiplicative depth. 

The energy efficiency of CiM-HE is due to two main factors: (i) CiM-HE does not use algorithmic optimizations like NTT or RNS, (ii) computation is performed in memory. The impact of these factors is explained below.
For (i), the \textit{PolyMult} primitive that relies on additions and shifts (Karatsuba) requires simpler hardware when compared to the hardware required for NTT in \cite{roy19_hpca}. The latter requires the design of large multiplier units and special modules for performing polynomial lift and scaling (which are not used/needed in our design). The modulo reduction circuitry is also more complicated than the bit shifters in CiM-HE, which employs moduli $q$ that are powers of 2.
For (ii), CiM avoids large amounts of data transfers to processing units while performing HE operations. CiM-HE takes advantage of data placement (as described in Sec. \ref{sec:CiM-HE_framework}) to perform Boolean logic at the bitline level with the use of customized sense amplifiers. Boolean operations can be leveraged to implement more complex functions, e.g. arithmetic additions, with a lower area overhead when compared to implementing arithmetic logic units (ALUs) from scratch \cite{reis19, aga17}.

Note that if one were to decide not to leverage optimizations such as NTT and RNS in a design that performs conventional data-processing, i.e., not in-memory, we expect item \textbf{(ii)} above to significantly influence the associated speedups/energy savings when compared to CiM-HE implementations. This is because it is not easy to design processing units that can match the processing power of CiM, which performs logic at the sense amplifier level and takes advantage from inherently high internal bandwidth of the memory \cite{jain17}.

\begin{table}[t]
\centering
\setlength\extrarowheight{2pt}
\caption{Parameters* used in HE Implementations}
\begin{tabular}{|l|c|c|c|}
\hline
\multicolumn{1}{|c|}{HE implementation} & log q & n & Optimizations \\ \hline
Bos, et al. \cite{bos2017privacy} & 186 & 4096 & Yes \\ \hline
Roy, et al. \cite{roy19_hpca} & 180 & 4096 & Yes \\ \hline
CiM-HE (this work) & 180 & 4096 & No \\ \hline
\end{tabular}
\label{tab:comparison_other_works}
\begin{flushright}
\scriptsize
\textbf{*Note:} These parameter settings enable a security level\\ of 80-bit and a multiplicative depth of 4 
\end{flushright}
\end{table}

\begin{table}[]
\centering
\setlength\extrarowheight{2pt}
\caption{Runtime and energy of \textit{HomMult} running on different HE platforms}
\begin{tabular}{|c|c|c|c|c|c|}
\hline
\begin{tabular}[c]{@{}c@{}}Figure of\\ Merit\\ (HomMult)\end{tabular} & \begin{tabular}[c]{@{}c@{}}Bos, et al.\\ \cite{bos2017privacy}\end{tabular} & \textbf{Imp.} & \begin{tabular}[c]{@{}c@{}}Roy, et al. \\ \cite{roy19_hpca}\end{tabular} & \textbf{Imp.} & CiM-HE \\ \hline
Runtime & 33 ms & \textbf{14.3x} & 5 ms & \textbf{2.2x} & 2.3 ms \\ \hline
Energy & 1.3 J & \textbf{2632.4x} & 43.5 mJ & \textbf{88.1x} & 494.0 $\mu$J \\ \hline
\end{tabular}
\label{tab:comparison_other_works2}
\end{table}

\section{Conclusion and Future Work}
\label{sec:conclusion}
We propose a CiM architecture that realizes essential operations for the B/FV scheme, a well-known SHE scheme. Our CiM-HE architecture consists of customized CMOS peripherals such as sense amplifiers, adders, bit shifters, and sequencing circuits. The peripherals 
and memory cells are based on a 14nm FinFET technology. Circuit-level evaluation of the CiM-HE design indicates maximum (minimum) speedups of 9.1x (4.6x) and maximum (minimum) energy savings of 266.4x (532.8x) for homomorphic multiplications (the most expensive HE operation). Furthermore, we evaluate arithmetic mean, variance, linear regression, and inference tasks using CiM-HE. The speedups (and energy savings) are associated with the dominant HE operation required by each task. Furthermore, our results support the idea that using multiple CiM banks can improve CiM-HE speedups 
by allowing them to operate on different ciphertexts in parallel, taking advantage of internal memory bandwidth. However, a more efficient multi-bank approach for CiM-HE may require larger memory sizes. Therefore, we plan to study the use of CiM-HE in the main memory as an alternative to the cache by employing denser memory cells based on CMOS (DRAM) or emerging technologies. We also plan to integrate algorithmic optimization techniques into our design to further increase the speedup of CiM-HE.


\bibliographystyle{./IEEEtran.bst}
\bibliography{references}

\begin{thebibliography}{10}
\providecommand{\url}[1]{#1}
\csname url@samestyle\endcsname
\providecommand{\newblock}{\relax}
\providecommand{\bibinfo}[2]{#2}
\providecommand{\BIBentrySTDinterwordspacing}{\spaceskip=0pt\relax}
\providecommand{\BIBentryALTinterwordstretchfactor}{4}
\providecommand{\BIBentryALTinterwordspacing}{\spaceskip=\fontdimen2\font plus
\BIBentryALTinterwordstretchfactor\fontdimen3\font minus
  \fontdimen4\font\relax}
\providecommand{\BIBforeignlanguage}[2]{{%
\expandafter\ifx\csname l@#1\endcsname\relax
\typeout{** WARNING: IEEEtran.bst: No hyphenation pattern has been}%
\typeout{** loaded for the language `#1'. Using the pattern for}%
\typeout{** the default language instead.}%
\else
\language=\csname l@#1\endcsname
\fi
#2}}
\providecommand{\BIBdecl}{\relax}
\BIBdecl

\bibitem{gentry09_FHE}
C.~Gentry, ``{Fully Homomorphic Encryption Using Ideal Lattices},'' in
  \emph{Proceedings of the 41st Annual ACM Symposium on Theory of Computing},
  ser. STOC '09.\hskip 1em plus 0.5em minus 0.4em\relax New York, NY, USA: ACM,
  2009, pp. 169--178. [Online]. Available:
  \url{http://doi.acm.org/10.1145/1536414.1536440}

\bibitem{ducas15_fhew}
L.~Ducas and D.~Micciancio, ``{FHEW: bootstrapping homomorphic encryption in
  less than a second},'' in \emph{Eurocrypt}.\hskip 1em plus 0.5em minus
  0.4em\relax Springer, 2015, pp. 617--640.

\bibitem{fan12_BFV}
J.~Fan and F.~Vercauteren, ``{Somewhat Practical Fully Homomorphic
  Encryption},'' \emph{IACR Cryptology ePrint Archive}, 2012.

\bibitem{bos2013improved}
J.~W. Bos, K.~Lauter, J.~Loftus, and M.~Naehrig, ``Improved security for a
  ring-based fully homomorphic encryption scheme,'' in \emph{IMA International
  Conference on Cryptography and Coding}.\hskip 1em plus 0.5em minus
  0.4em\relax Springer, 2013, pp. 45--64.

\bibitem{roy19_hpca}
S.~{Sinha Roy}, F.~{Turan}, K.~{Jarvinen}, F.~{Vercauteren}, and
  I.~{Verbauwhede}, ``{FPGA-Based High-Performance Parallel Architecture for
  Homomorphic Computing on Encrypted Data},'' in \emph{HPCA}, Feb 2019, pp.
  387--398.

\bibitem{jeloka16}
S.~{Jeloka}, N.~B. {Akesh}, D.~{Sylvester}, and D.~{Blaauw}, ``{A 28 nm
  Configurable Memory (TCAM/BCAM/SRAM) Using Push-Rule 6T Bit Cell Enabling
  Logic-in-Memory},'' \emph{IEEE Journal of Solid-State Circuits}, vol.~51, pp.
  1009--1021, April 2016.

\bibitem{pawlowski2011hybrid}
J.~T. Pawlowski, ``{Hybrid memory cube (HMC)},'' in \emph{IEEE HCS}.\hskip 1em
  plus 0.5em minus 0.4em\relax IEEE, 2011, pp. 1--24.

\bibitem{ahn15_graph}
J.~{Ahn}, S.~{Hong}, S.~{Yoo}, O.~{Mutlu}, and K.~{Choi}, ``{A scalable
  processing-in-memory accelerator for parallel graph processing},'' in
  \emph{ISCA}, June 2015, pp. 105--117.

\bibitem{ahn15_pimenabled}
J.~{Ahn}, S.~{Yoo}, O.~{Mutlu}, and K.~{Choi}, ``{PIM-enabled instructions: A
  low-overhead, locality-aware processing-in-memory architecture},'' in
  \emph{ISCA}, June 2015, pp. 336--348.

\bibitem{glova19}
A.~O. {Glova}, I.~{Akgun}, S.~{Li}, X.~{Hu}, and Y.~{Xie}, ``{Near-Data
  Acceleration of Privacy-Preserving Biomarker Search with 3D-Stacked
  Memory},'' in \emph{DATE}, March 2019, pp. 800--805.

\bibitem{jain17}
S.~Jain, A.~Ranjan, K.~Roy, and A.~Raghunathan, ``{Computing in Memory With
  Spin-Transfer Torque Magnetic RAM},'' \emph{TVLSI}, vol.~PP, pp. 1--14, 2017.

\bibitem{aga17}
S.~{Aga}, S.~{Jeloka}, A.~{Subramaniyan}, S.~{Narayanasamy}, D.~{Blaauw}, and
  R.~{Das}, ``Compute caches,'' in \emph{HPCA}, Feb 2017, pp. 481--492.

\bibitem{imani19_floatpim}
M.~Imani, S.~Gupta, Y.~Kim, and T.~Rosing, ``{FloatPIM: In-memory Acceleration
  of Deep Neural Network Training with High Precision}, booktitle =
  {ISCA}.''\hskip 1em plus 0.5em minus 0.4em\relax New York, NY, USA: ACM,
  2019, pp. 802--815.

\bibitem{chi19_prime}
P.~{Chi}, S.~{Li}, C.~{Xu}, T.~{Zhang}, J.~{Zhao}, Y.~{Liu}, Y.~{Wang}, and
  Y.~{Xie}, ``{PRIME: A Novel Processing-in-Memory Architecture for Neural
  Network Computation in ReRAM-Based Main Memory},'' in \emph{ISCA}, June 2016,
  pp. 27--39.

\bibitem{feinberg_memristive}
B.~{Feinberg}, U.~K.~R. {Vengalam}, N.~{Whitehair}, S.~{Wang}, and E.~{Ipek},
  ``{Enabling Scientific Computing on Memristive Accelerators},'' in \emph{2018
  ACM/IEEE 45th Annual International Symposium on Computer Architecture
  (ISCA)}, 2018, pp. 367--382.

\bibitem{li16}
S.~Li, C.~Xu, Q.~Zou, J.~Zhao, Y.~Lu, and Y.~Xie, ``{Pinatubo: A
  processing-in-memory architecture for bulk bitwise operations in emerging
  non-volatile memories},'' in \emph{DAC}, 2016, pp. 1--6.

\bibitem{reis18}
D.~Reis, M.~Niemier, and X.~S. Hu, ``{Computing in Memory with FeFETs},'' in
  \emph{ISLPED}.\hskip 1em plus 0.5em minus 0.4em\relax New York, NY, USA: ACM,
  2018, pp. 24:1--24:6. [Online]. Available:
  \url{http://doi.acm.org/10.1145/3218603.3218640}

\bibitem{laguna19_fewshot}
A.~F. Laguna, X.~Yin, D.~Reis, M.~Niemier, and X.~S. Hu, ``{Ferroelectric FET
  Based In-Memory Computing for Few-Shot Learning},'' in \emph{GLSVLSI}.\hskip
  1em plus 0.5em minus 0.4em\relax New York, NY, USA: ACM, 2019, pp. 373--378.
  [Online]. Available: \url{http://doi.acm.org/10.1145/3299874.3319450}

\bibitem{gupta19}
S.~{Gupta}, M.~{Imani}, H.~{Kaur}, and T.~S. {Rosing}, ``{NNPIM: A Processing
  In-Memory Architecture for Neural Network Acceleration},'' \emph{IEEE
  Transactions on Computers}, vol.~68, pp. 1325--1337, Sep. 2019.

\bibitem{brakerski2014leveled}
Z.~Brakerski, C.~Gentry, and V.~Vaikuntanathan, ``{Leveled) fully homomorphic
  encryption without bootstrapping},'' \emph{ACM Transactions on Computation
  Theory (TOCT)}, vol.~6, pp. 1--36, 2014.

\bibitem{GSW}
C.~Gentry, A.~Sahai, and B.~Waters, ``{Homomorphic encryption from learning
  with errors: Conceptually-simpler, asymptotically-faster, attribute-based},''
  in \emph{Annual Cryptology Conference}.\hskip 1em plus 0.5em minus
  0.4em\relax Springer, 2013, pp. 75--92.

\bibitem{TFHE}
I.~Chillotti, N.~Gama, M.~Georgieva, and M.~Izabach{\`e}ne, ``{TFHE}: Fast
  fully homomorphic encryption library,'' August 2016,
  https://tfhe.github.io/tfhe/.

\bibitem{cheon17_CKKS}
J.~H. Cheon, A.~Kim, M.~Kim, and Y.~Song, ``Homomorphic encryption for
  arithmetic of approximate numbers,'' in \emph{Eurocrypt}.\hskip 1em plus
  0.5em minus 0.4em\relax Springer, 2017, pp. 409--437.

\bibitem{duarte15_bsim_cmg}
J.~P. {Duarte}, S.~{Khandelwal}, A.~{Medury}, C.~{Hu}, P.~{Kushwaha},
  H.~{Agarwal}, A.~{Dasgupta}, and Y.~S. {Chauhan}, ``{BSIM-CMG: Standard
  FinFET compact model for advanced circuit design},'' in \emph{ESSCIRC}, Sep.
  2015, pp. 196--201.

\bibitem{chen17_SEAL}
H.~Chen, K.~Laine, and R.~Player, ``{Simple encrypted arithmetic library-SEAL
  v2. 1},'' in \emph{Eurocrypt}.\hskip 1em plus 0.5em minus 0.4em\relax
  Springer, 2017, pp. 3--18.

\bibitem{bos2017privacy}
J.~W. Bos, W.~Castryck, I.~Iliashenko, and F.~Vercauteren, ``{Privacy-friendly
  forecasting for the smart grid using homomorphic encryption and the group
  method of data handling},'' in \emph{International Conference on Cryptology
  in Africa}.\hskip 1em plus 0.5em minus 0.4em\relax Springer, 2017, pp.
  184--201.

\bibitem{brakerski2013classical}
Z.~Brakerski, A.~Langlois, C.~Peikert, O.~Regev, and D.~Stehl{\'e},
  ``{Classical hardness of learning with errors},'' in \emph{Proceedings of the
  forty-fifth annual ACM symposium on Theory of computing}.\hskip 1em plus
  0.5em minus 0.4em\relax ACM, 2013, pp. 575--584.

\bibitem{halevi2013design}
S.~Halevi and V.~Shoup, ``{Design and implementation of a
  homomorphic-encryption library},'' \emph{IBM Research (Manuscript)}, vol.~6,
  pp. 12--15, 2013.

\bibitem{riazi2020heax}
M.~S. Riazi, K.~Laine, B.~Pelton, and W.~Dai, ``{HEAX: An Architecture for
  Computing on Encrypted Data},'' in \emph{Proceedings of the Twenty-Fifth
  International Conference on Architectural Support for Programming Languages
  and Operating Systems}, 2020, pp. 1295--1309.

\bibitem{cousins17_fpga_accelerator}
D.~B. {Cousins}, K.~{Rohloff}, and D.~{Sumorok}, ``{Designing an
  FPGA-Accelerated Homomorphic Encryption Co-Processor},'' \emph{IEEE
  Transactions on Emerging Topics in Computing}, vol.~5, pp. 193--206, April
  2017.

\bibitem{bajard2016full}
J.-C. Bajard, J.~Eynard, M.~A. Hasan, and V.~Zucca, ``{A full RNS variant of FV
  like somewhat homomorphic encryption schemes},'' in \emph{International
  Conference on Selected Areas in Cryptography}.\hskip 1em plus 0.5em minus
  0.4em\relax Springer, 2016, pp. 423--442.

\bibitem{halevi2019improved}
S.~Halevi, Y.~Polyakov, and V.~Shoup, ``{An improved RNS variant of the BFV
  homomorphic encryption scheme},'' in \emph{Cryptographers’ Track at the RSA
  Conference}.\hskip 1em plus 0.5em minus 0.4em\relax Springer, 2019, pp.
  83--105.

\bibitem{poppelmann2015accelerating}
T.~P{\"o}ppelmann, M.~Naehrig, A.~Putnam, and A.~Macias, ``Accelerating
  homomorphic evaluation on reconfigurable hardware,'' in \emph{International
  Workshop on Cryptographic Hardware and Embedded Systems}.\hskip 1em plus
  0.5em minus 0.4em\relax Springer, 2015, pp. 143--163.

\bibitem{cheon2018bootstrapping}
J.~H. Cheon, K.~Han, A.~Kim, M.~Kim, and Y.~Song, ``{Bootstrapping for
  approximate homomorphic encryption},'' in \emph{Annual International
  Conference on the Theory and Applications of Cryptographic Techniques}.\hskip
  1em plus 0.5em minus 0.4em\relax Springer, 2018, pp. 360--384.

\bibitem{halevi2015bootstrapping}
S.~Halevi and V.~Shoup, ``Bootstrapping for helib,'' in \emph{Annual
  International conference on the theory and applications of cryptographic
  techniques}.\hskip 1em plus 0.5em minus 0.4em\relax Springer, 2015, pp.
  641--670.

\bibitem{wang14}
Y.~{Wang}, H.~{Yu}, D.~{Sylvester}, and P.~{Kong}, ``{Energy efficient
  in-memory AES encryption based on nonvolatile domain-wall nanowire},'' in
  \emph{DATE}, March 2014, pp. 1--4.

\bibitem{xie18}
M.~{Xie}, S.~{Li}, A.~O. {Glova}, J.~{Hu}, Y.~{Wang}, and Y.~{Xie}, ``{AIM:
  Fast and energy-efficient AES in-memory implementation for emerging
  non-volatile main memory},'' in \emph{DATE}, March 2018, pp. 625--628.

\bibitem{bian17_SCAM}
S.~{Bian}, M.~{Hiromoto}, and T.~{Sato}, ``{SCAM: Secured content addressable
  memory based on homomorphic encryption},'' in \emph{DATE}, March 2017, pp.
  984--989.

\bibitem{ducas2015fhew}
L.~Ducas and D.~Micciancio, ``{FHEW: bootstrapping homomorphic encryption in
  less than a second},'' in \emph{Eurocrypt}.\hskip 1em plus 0.5em minus
  0.4em\relax Springer, 2015, pp. 617--640.

\bibitem{reis19}
D.~{Reis}, M.~{Niemier}, and X.~S. {Hu}, ``{A Computing-in-Memory Engine for
  Searching on Homomorphically Encrypted Data},'' \emph{IEEE Journal on
  Exploratory Solid-State Computational Devices and Circuits}, pp. 1--1, 2019.

\bibitem{roy15_modular}
S.~Sinha~Roy, K.~J{\"a}rvinen, F.~Vercauteren, V.~Dimitrov, and I.~Verbauwhede,
  ``{Modular Hardware Architecture for Somewhat Homomorphic Function
  Evaluation},'' in \emph{CHES 2015}, T.~G{\"u}neysu and H.~Handschuh,
  Eds.\hskip 1em plus 0.5em minus 0.4em\relax Berlin, Heidelberg: Springer
  Berlin Heidelberg, pp. 164--184.

\bibitem{wang14_tvlsi}
W.~{Wang}, X.~{Huang}, N.~{Emmart}, and C.~{Weems}, ``{VLSI Design of a
  Large-Number Multiplier for Fully Homomorphic Encryption},'' \emph{TVLSI},
  vol.~22, pp. 1879--1887, 2014.

\bibitem{xiaolin14}
X.~Cao, C.~Moore, M.~O’Neill, N.~Hanley, and E.~O'Sullivan, ``{High-Speed
  Fully Homomorphic Encryption Over the Integers},'' 03 2014, pp. 169--180.

\bibitem{cilardo16_fpga}
A.~{Cilardo} and D.~{Argenziano}, ``{Securing the cloud with reconfigurable
  computing: An FPGA accelerator for homomorphic encryption},'' in \emph{DATE},
  March 2016, pp. 1622--1627.

\bibitem{albrecht2016subfield}
M.~Albrecht, S.~Bai, and L.~Ducas, ``{A subfield lattice attack on
  overstretched NTRU assumptions},'' in \emph{Annual International Cryptology
  Conference}.\hskip 1em plus 0.5em minus 0.4em\relax Springer, 2016, pp.
  153--178.

\bibitem{langlois2012hardness}
A.~Langlois and D.~Stehl{\'e}, ``{Hardness of decision (R) LWE for any
  modulus},'' Citeseer, Tech. Rep., 2012.

\bibitem{peikert2009public}
C.~Peikert, ``Public-key cryptosystems from the worst-case shortest vector
  problem,'' in \emph{STOC}.\hskip 1em plus 0.5em minus 0.4em\relax ACM, 2009,
  pp. 333--342.

\bibitem{regev2009lattices}
O.~Regev, ``{On lattices, learning with errors, random linear codes, and
  cryptography},'' \emph{Journal of the ACM (JACM)}, vol.~56, p.~34, 2009.

\bibitem{lindner2011better}
R.~Lindner and C.~Peikert, ``{Better key sizes (and attacks) for LWE-based
  encryption},'' in \emph{Cryptographers’ Track at the RSA Conference}.\hskip
  1em plus 0.5em minus 0.4em\relax Springer, 2011, pp. 319--339.

\bibitem{karatsuba1962multiplication}
A.~A. Karatsuba and Y.~P. Ofman, ``{Multiplication of many-digital numbers by
  automatic computers},'' in \emph{Doklady Akademii Nauk}, vol. 145,
  no.~2.\hskip 1em plus 0.5em minus 0.4em\relax Russian Academy of Sciences,
  1962, pp. 293--294.

\bibitem{cesari1996performance}
G.~Cesari and R.~Maeder, ``{Performance analysis of the parallel Karatsuba
  multiplication algorithm for distributed memory architectures},''
  \emph{Journal of Symbolic Computation}, vol.~21, pp. 467--473, 1996.

\bibitem{barker2018transitioning}
E.~Barker and A.~Roginsky, ``{Transitioning the use of cryptographic algorithms
  and key lengths},'' National Institute of Standards and Technology, Tech.
  Rep., 2018.

\bibitem{albrecht2015concrete}
M.~R. Albrecht, R.~Player, and S.~Scott, ``{On the concrete hardness of
  learning with errors},'' \emph{Journal of Mathematical Cryptology}, vol.~9,
  pp. 169--203, 2015.

\bibitem{powerstat}
``{Powerstat, a power consumption calculator for Ubuntu},'' Online.
  \url{http://launchpad.net/ubuntu/xenial/+package/powerstat}, date:
  2019-09-08.

\bibitem{hspice2018inc}
{Synopsys Inc.}, ``{HSPICE},'' \emph{Version O-2018.09-1}, 2018.

\bibitem{martins15_stdcell}
M.~Martins, J.~M. Matos, R.~P. Ribas, A.~Reis, G.~Schlinker, L.~Rech, and
  J.~Michelsen, ``{Open Cell Library in 15Nm FreePDK Technology},'' in
  \emph{ISPD}.\hskip 1em plus 0.5em minus 0.4em\relax New York, NY, USA: ACM,
  2015, pp. 171--178. [Online]. Available:
  \url{http://doi.acm.org/10.1145/2717764.2717783}

\bibitem{gao2020eva}
D.~Gao, D.~Reis, X.~S. Hu, and C.~Zhuo, ``{Eva-CiM: A System-Level Performance
  and Energy Evaluation Framework for Computing-in-Memory Architectures},''
  \emph{IEEE Transactions on Computer-Aided Design of Integrated Circuits and
  Systems}, 2020.

\bibitem{dong12_nvsim}
X.~{Dong}, C.~{Xu}, Y.~{Xie}, and N.~P. {Jouppi}, ``{NVSim: A Circuit-Level
  Performance, Energy, and Area Model for Emerging Nonvolatile Memory},''
  \emph{TCAD}, vol.~31, pp. 994--1007, July 2012.

\bibitem{borkar2010exascale}
S.~Borkar, ``{The exascale challenge},'' in \emph{VLSI-DAT}.\hskip 1em plus
  0.5em minus 0.4em\relax IEEE, 2010, pp. 2--3.

\bibitem{bhanushali2015freepdk15}
K.~Bhanushali and W.~R. Davis, ``{FreePDK15: An open-source predictive process
  design kit for 15nm FinFET technology},'' in \emph{Proceedings of the 2015
  Symposium on International Symposium on Physical Design}, 2015, pp. 165--170.

\bibitem{deng2012mnist}
L.~Deng, ``{The mnist database of handwritten digit images for machine learning
  research [best of the web]},'' \emph{IEEE Signal Processing Magazine},
  vol.~29, pp. 141--142, 2012.

\bibitem{cryptoexperts}
``{CryptoExperts,``FV-NFLlib"},'' https://github.com/CryptoExperts/ FV-NFLlib,
  online. Year: 2016.

\bibitem{intelcpu}
``{Intel Kaby Lake Core i7-7700K, i7-7700, i5-7600K, i5-7600 Review.}''
  http://www.tomshardware.com/reviews/intel-kaby-lake-core-i7-7700k-i7-7700-i5-7600k-i5-7600,4870.html,
  online. Accessed on 11/2019.

\end{thebibliography}

\end{document}